\documentclass[]{mn2e}

\usepackage{graphicx}

\def\Earth{$\rm _{Earth}$}
\def\Jupiter{$\rm _{Jupiter}$}

\title[Methods for exomoon characterisation]{Methods for exomoon characterisation:
{combining} transit photometry and the Rossiter-McLaughlin effect}  
\author[A. E. Simon,  Gy. M. Szab\'o, 
K. Szatm\'ary and L. L. Kiss]{
A. E. Simon$^{1,2}$\thanks{E-mail: atthys@konkoly.hu},
Gy. M. Szab\'o$^{1,2}$\thanks{E-mail: szgy@konkoly.hu},
K. Szatm\'ary$^{2}$\thanks{E-mail: k.szatmary@physx.u-szeged.hu}
and L. L. Kiss$^{1,3}$\thanks{E-mail: kiss@konkoly.hu}\\
$^1$Konkoly Observatory of the Hungarian Academy of Sciences, PO. Box 67,
H-1525 Budapest, Hungary\\
$^2$Department of Experimental Physics and Astronomical Observatory, University of Szeged, 6720 Szeged, Hungary\\
$^3$Sydney Institute for Astronomy, School of Physics A28, University of Sydney,
NSW 2006, Australia
}

\begin{document}

\date{Accepted  Received  in original form }


\maketitle

\label{firstpage}

\begin{abstract}
It has been suggested that moons around transiting exoplanets may cause observable signal in transit
photometry or in the Rossiter-McLaughlin (RM) effect.
In this paper a detailed analysis of parameter reconstruction
from the RM effect is presented for various planet-moon configurations, described with 20 parameters.
We also demonstrate the benefits of combining photometry with the RM effect.
We simulated 2.7$\times 10^9$ configurations of a generic transiting
system to map the confidence region of the parameters of the moon, 
find the correlated parameters and determine the validity of reconstructions.
The main conclusion is that the strictest constraints  from the RM effect 
are expected for the radius of the moon.
In some cases there is also meaningful information on its orbital period.
When the transit time of the moon is exactly known, for example, from transit photometry, the angle parameters of the moon's orbit will also be constrained from the RM effect.
From transit light curves the mass can be determined, and combining this result
with the radius from the RM effect, the experimental determination of the density 
of the moon is also possible.
\end{abstract}


\section{Introduction}

The number of known transiting exoplanets is rapidly increasing,
which has recently inspired significant interest as to whether 
they can host a detectable moon
(e.g. Szab\'o et al. 2006, Simon et al. 2007, Kipping 2008, 2009,
Simon et al. 2009, Kipping et al. 2009).
Historically, our Moon has constantly inspired scientific research and it has
played a key role in supporting life on Earth (e.g.
Wagner 1936, Asimov 1979). It may be that the 
presence of a large
exomoon is a {\it sine qua non} requirement for the development of an
intelligent civilization on an exoplanet.

Although there has been no such example where the presence of a satellite was
proven, several methods have already been investigated for such a detection
in the future (barycentric Transit Timing Variation, TTV
Sartoretti \&{} Schneider 1999,
Kipping, 2008; photocentric Transit Timing Variation, TTV$_{\rm p}$
Szab\'o et al. 2006, Simon et al. 2007; Transit
Duration Variation, TDV, Kipping, 2009; Time-of-Arrival 
analysis of pulsars, Lewis et al. 2008; microlensing, Liebig \& Wambsganss 2009).
Deviations from perfect periodic timing of 
transits might suggest the presence of a 
moon (D\'\i{}az et al. 2008),
perturbing planets (Agol et al. 2005) or indicate
periastron precession (P\'al \&{} Kocsis 2008).

All these methods (excluding microlensing) rely on transit photometry. 
In the era of ultraprecise space photometry (CoRoT, Kepler), 
one can expect accurate light curves to 0.1 mmag  that
promises the discovery of Moon-like satellites of Earth-like planets 
(Szab\'o et al. 2006, Kipping et. al 2009). Additionally, 
the $\sim$1 cm/s velocimetric accuracy is promised with laser frequency 
combs (Li et al. 2008). Radial velocity measurements during
a transit already play an important role in understanding
the planet via its Rossiter-McLaughlin (RM) effect (Gaudi \&{} Winn 2007),
and the prospects of exomoon detection in this way are quite encouraging.


Here we continue our previous investigations by invoking the RM effect
as a possible tool in characterising exomoons.
Earlier we described a photometric method,
the Photocentric Transit Timing Variation, TTV$_{\rm p}$ 
(Szab\'o{} et al. 2006)
that is
very sensitive to the presence of transiting moons. In Simon et al. (2007) we
examined which moons can be detected with that method in
space observatory measurements, with respect to different
values of M masses, R radii, and P orbital periods. 
In Simon et al. (2009) we demonstrated that
another method, based on the Rossiter-McLaughlin effect of the moon, is also
capable of attaining observational signature of a possible satellite. 
Now we give a full description of the parameter reconstruction from the RM effect.
An error analysis is also presented:
from simulated transits we examine which parameters of the 
satellite can be recovered at certain S/N of the measurements. 



\section{Simulations}

\begin{figure*}
  \noindent\includegraphics[angle=270,width=0.5\textwidth]{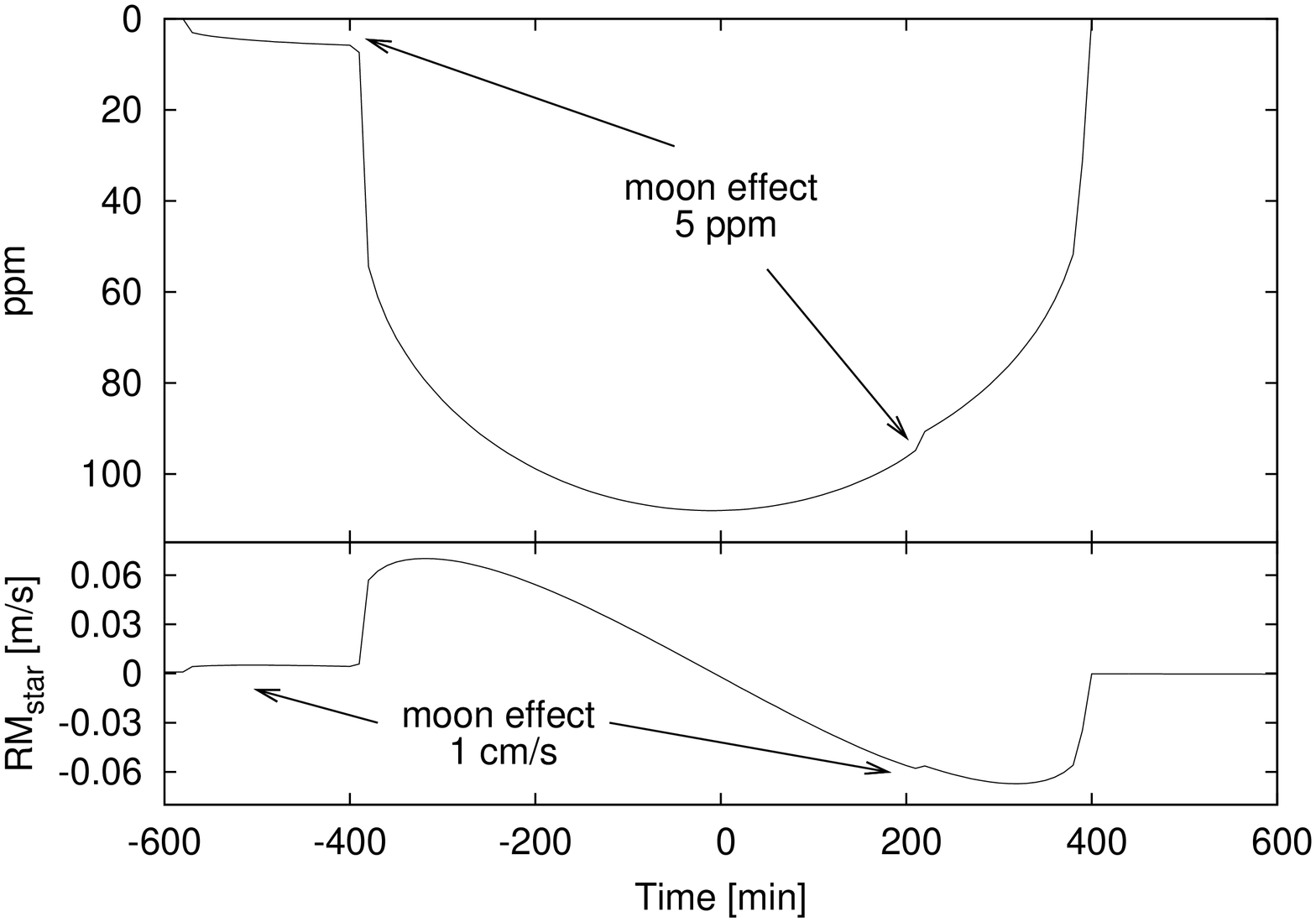}%
  \noindent\includegraphics[angle=270,width=0.5\textwidth]{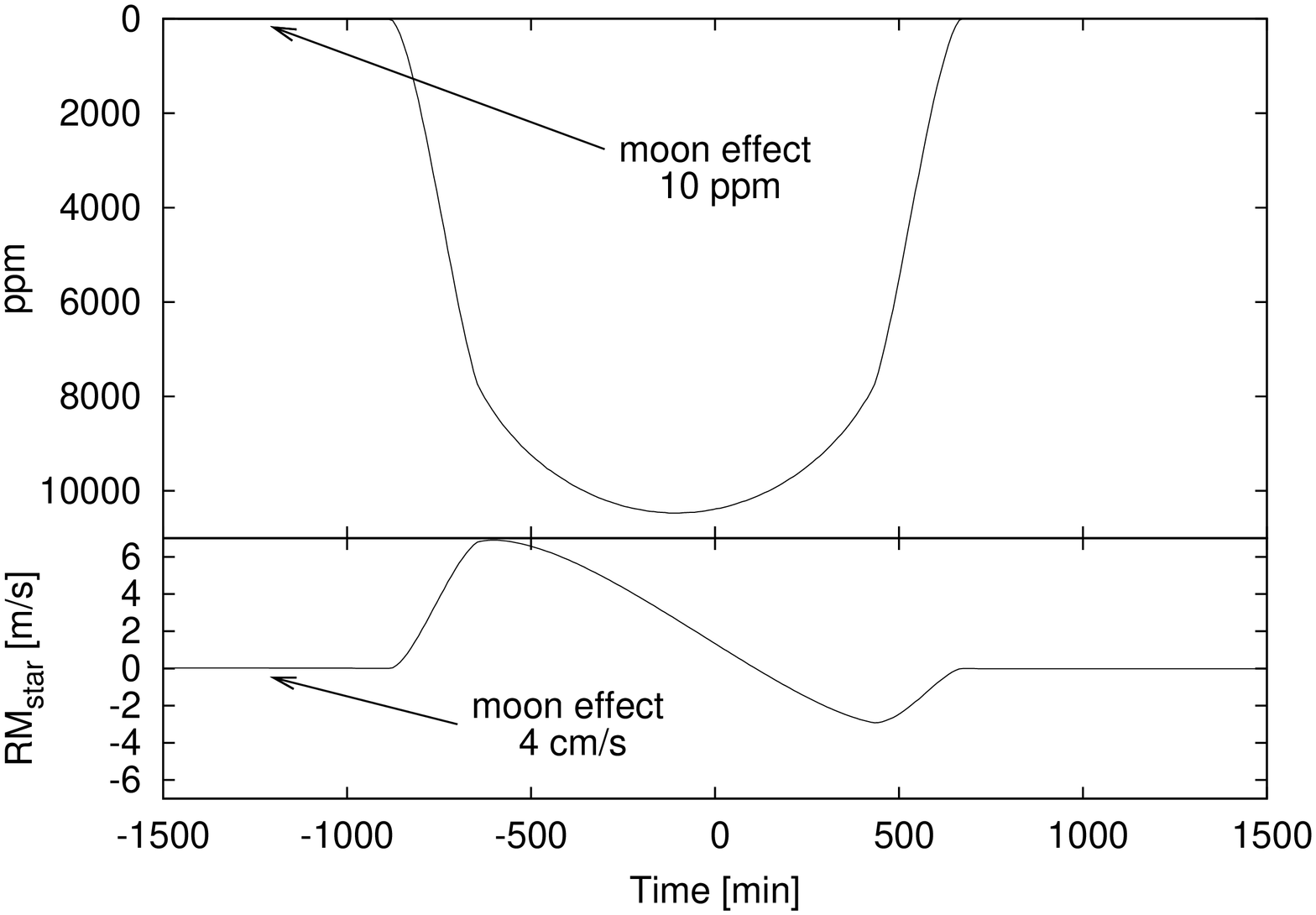}%
\vskip0.4cm
  \noindent\includegraphics[angle=270,width=0.5\textwidth]{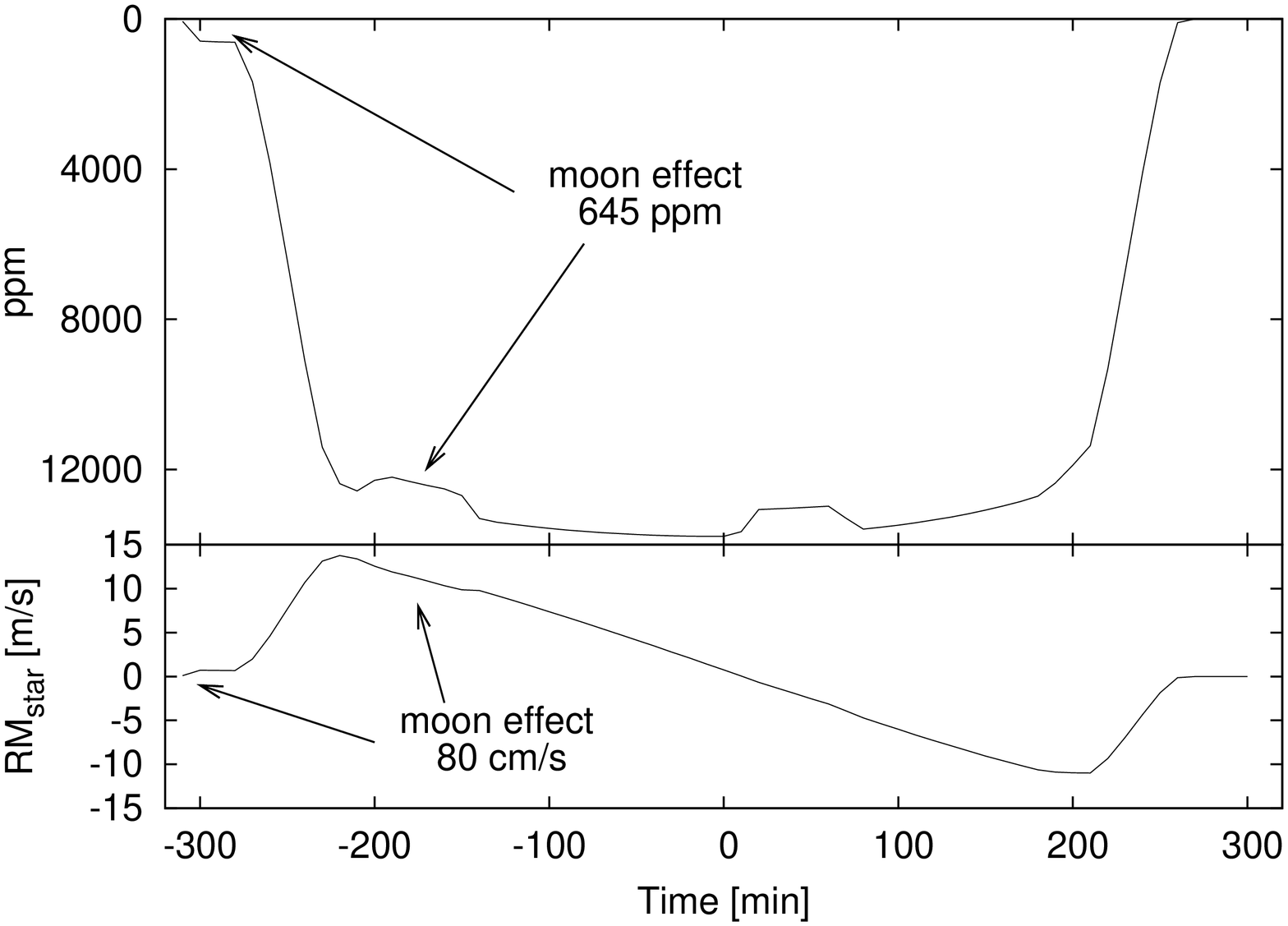}%
  \noindent\includegraphics[angle=270,width=0.5\textwidth]{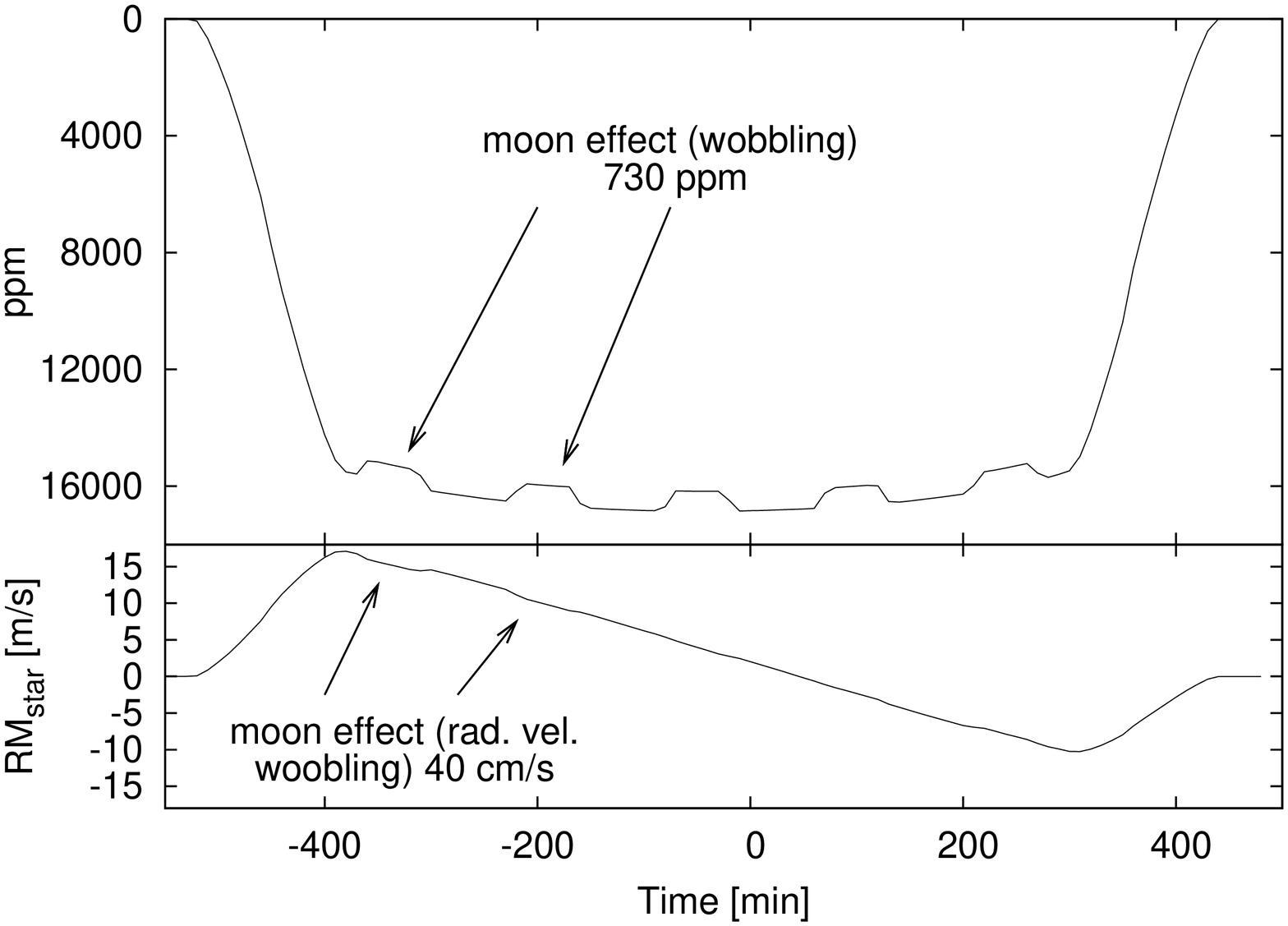}%
\vskip1cm
\caption{Sample transit simulations. The four panels show the
the light curves in the upper part, and the RM curves below.
Top left: the Earth-Moon system; top right: the Jupiter-Ganymede--like system;
bottom left: Simulation 3; bottom right: Simulation 4. See Table 1
for the parameters.}
\label{fig:1}       
\end{figure*}

\begin{table}
\caption{Input data of the simulations. `asc. node' means the angle between the line of sight and the line of intersection of the planet's orbital plane and the star's equatorial plane, measured in the star's equatorial plane.}
\label{tab:1}
\begin{tabular}{@{}lccc}
\hline\noalign{\smallskip}
{\bf Simulation 1}  \\
{\bf (``Earth")}  & {\bf Star}& {\bf Planet} & {\bf Satellite} \\
limb dark. $(u)$	&0.65	&	&	\\
mass $(M)$  & 1.00   M$_\odot$   &0.0032  M\Jupiter   &0.0123 M\Earth\\
radius $(R)$&   1.00 R$_\odot$   &  0.0920  R\Jupiter &  0.2720 R\Earth\\
rot. period $(P_{rot})$ &   28.00 days   & &\\
orb. period $(P)$ & & 365.25 days  & 27.30 days\\
inclination$(\iota)$  & &     90$^\circ$&    90$^\circ$\\
asc. node $(\Omega)$  & &     0$^\circ$ &   0$^\circ$\\
\hline\noalign{\smallskip}
{\bf Simulation 2}  \\
{\bf (``Jupiter")}  & {\bf Star}& {\bf Planet} & {\bf Satellite} \\
limb dark. $(u)$	&0.65	&	&	\\
mass  $(M)$  &1.00  M$_\odot$  &1.00 M\Jupiter    &0.0246 M\Earth\\
radius $(R)$ &  1.00 R$_\odot$ &  1.00 R\Jupiter  &  0.4125 R\Earth \\
rot. period $(P_{rot})$   &  28.00 days & &\\
orb. period $(P)$ &  & 4332.71 days &  7.15 days\\
inclination$(\iota)$    & &   70$^\circ$&    90$^\circ$\\
asc. node $(\Omega)$   & &    $-0.10^\circ$&    0$^\circ$\\
\hline\noalign{\smallskip}
{\bf Simulation 3}  & {\bf Star}& {\bf Planet} & {\bf Satellite} \\
limb dark. $(u)$	&0.20	&	&	\\
mass $(M)$ & 0.30   M$_\odot$& 0.15 M\Jupiter & 1 M\Earth \\
radius $(R)$ & 0.36   R$_\odot$& 0.40 R\Jupiter & 1 R\Earth \\
rot. period $(P_{rot})$ & 10 days & &\\
orb. period $(P)$ & & 600 days &0.3 days\\
inclination $(\iota)$& & 65$^\circ$ &80$^\circ$ \\
asc. node $(\Omega)$& & 0.04$^\circ$&  0$^\circ$ \\
 \hline\noalign{\smallskip}
{\bf Simulation 4}  & {\bf Star}& {\bf Planet} & {\bf Satellite} \\
limb dark. $(u)$	&0.20	&	&	\\
 mass $(M)$ &0.30   M$_\odot$&0.15 M\Jupiter &1 M\Earth \\
 radius $(R)$ &0.36   R$_\odot$&0.45 R\Jupiter &1 R\Earth \\
 rot. period $(P_{rot})$ &10 days& &\\
 orb. period $(P)$ & &4300 days &0.2 days \\
 inclination $(\iota)$ &&70$^\circ$ &80$^\circ$\\
 asc. node $(\Omega)$ && 0.04$^\circ$& 0$^\circ$ \\
\hline\noalign{\smallskip}
{\bf Simulation 5}  & {\bf Star}& {\bf Planet} & {\bf Satellite} \\
limb dark. $(u)$	&0.65	&	&	\\
 mass $(M)$ &0.80   M$_\odot$&0.20 M\Jupiter &1 M\Earth \\
 radius $(R)$ &0.83   R$_\odot$&0.40 R\Jupiter &1 R\Earth \\
 rot. period $(P_{rot})$ &3 days& &\\
 orb. period $(P)$ & & 200 days &5 days \\
 inclination $(\iota)$ &&75$^\circ$ &60$^\circ$\\
 asc. node $(\Omega)$ && 0.5$^\circ$& 45$^\circ$ \\
\hline
\end{tabular}
\end{table}

\begin{figure*}
  \hskip1.2cm\includegraphics[bb=114 60 706 480,width=0.4\textwidth]{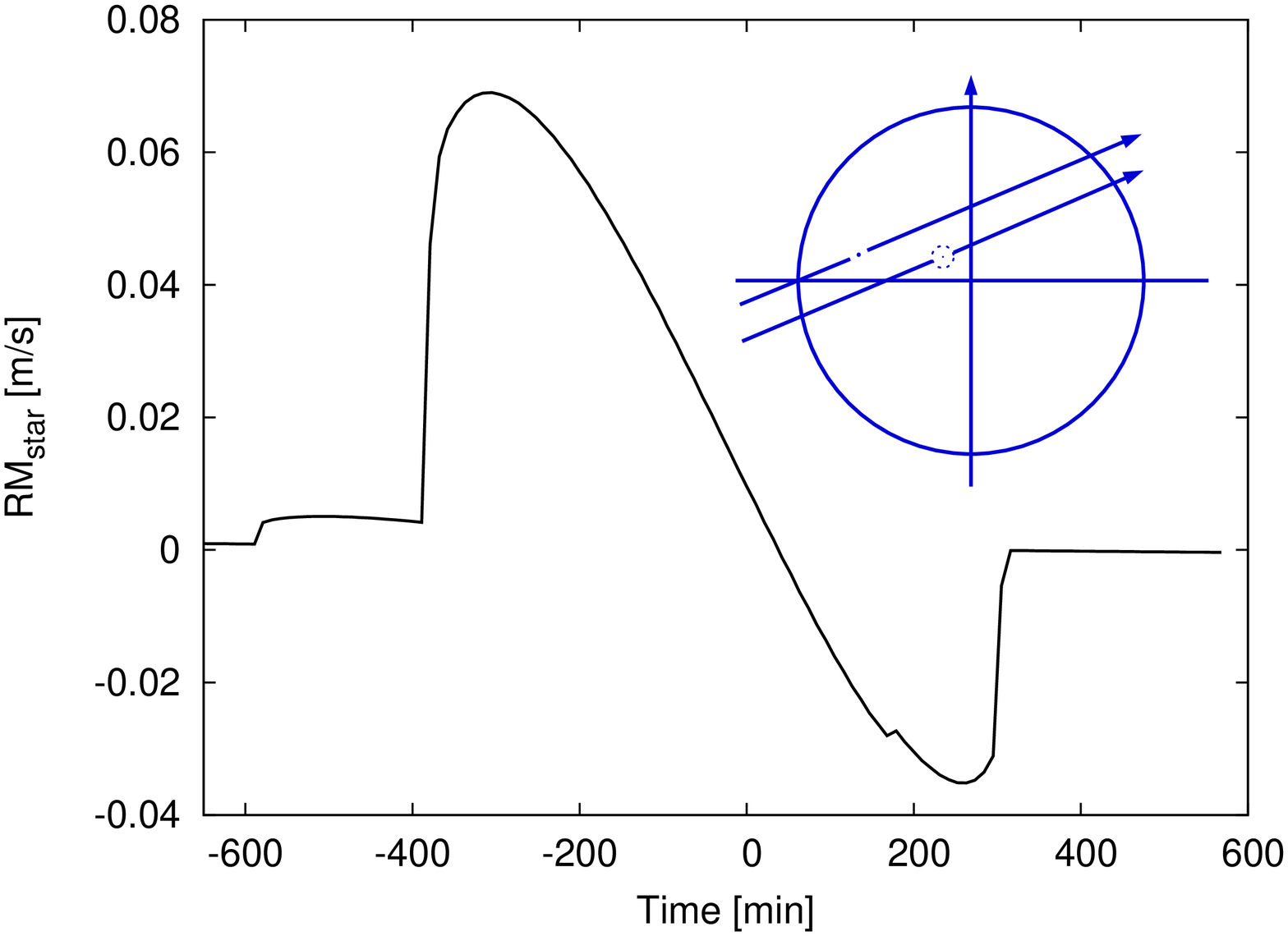}\hfill
  \includegraphics[bb=101 60 720 480,width=0.4\textwidth]{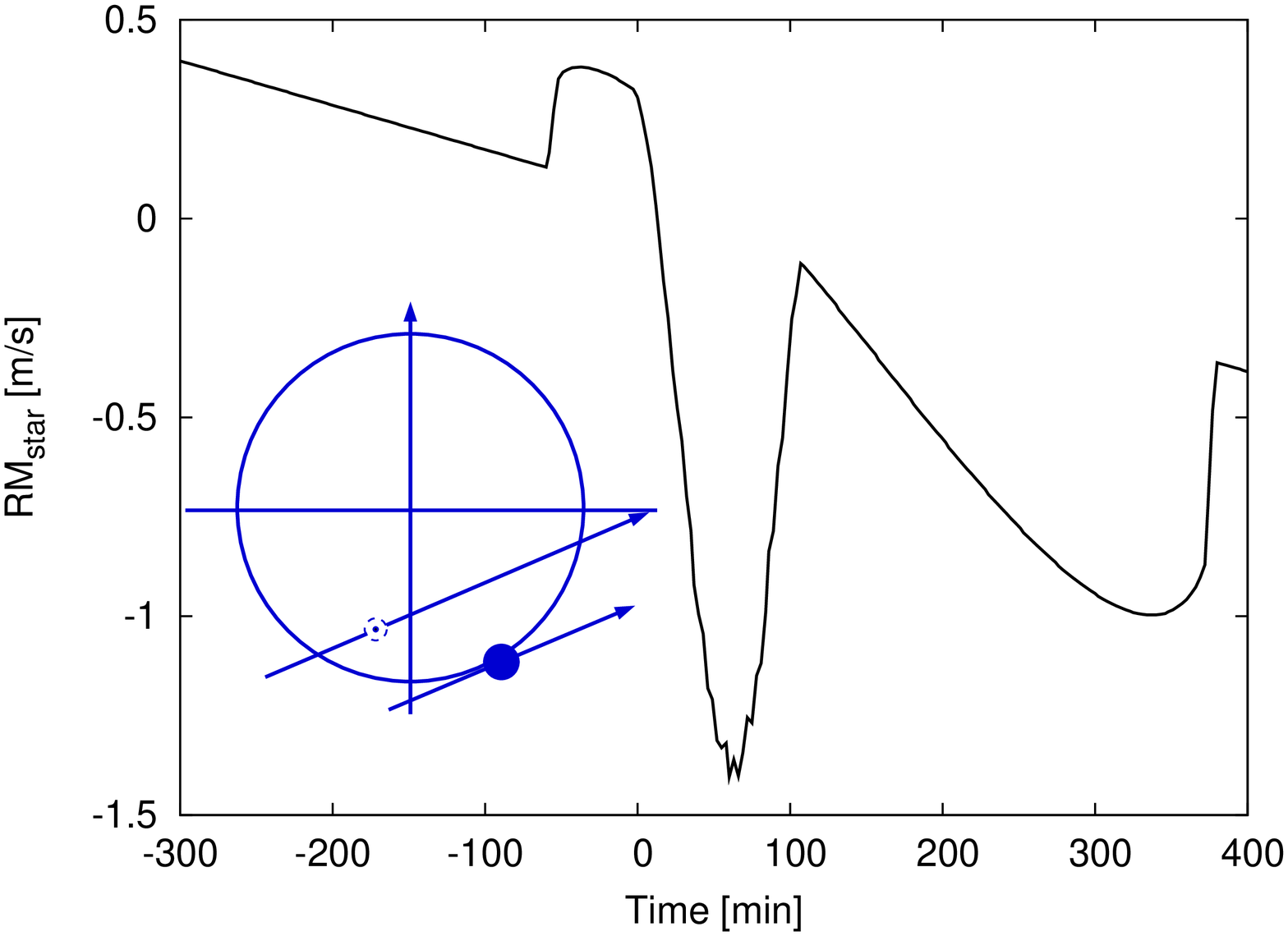}
\vskip1.4cm
  \hskip1.2cm\includegraphics[bb=90 60 706 480,width=0.4\textwidth]{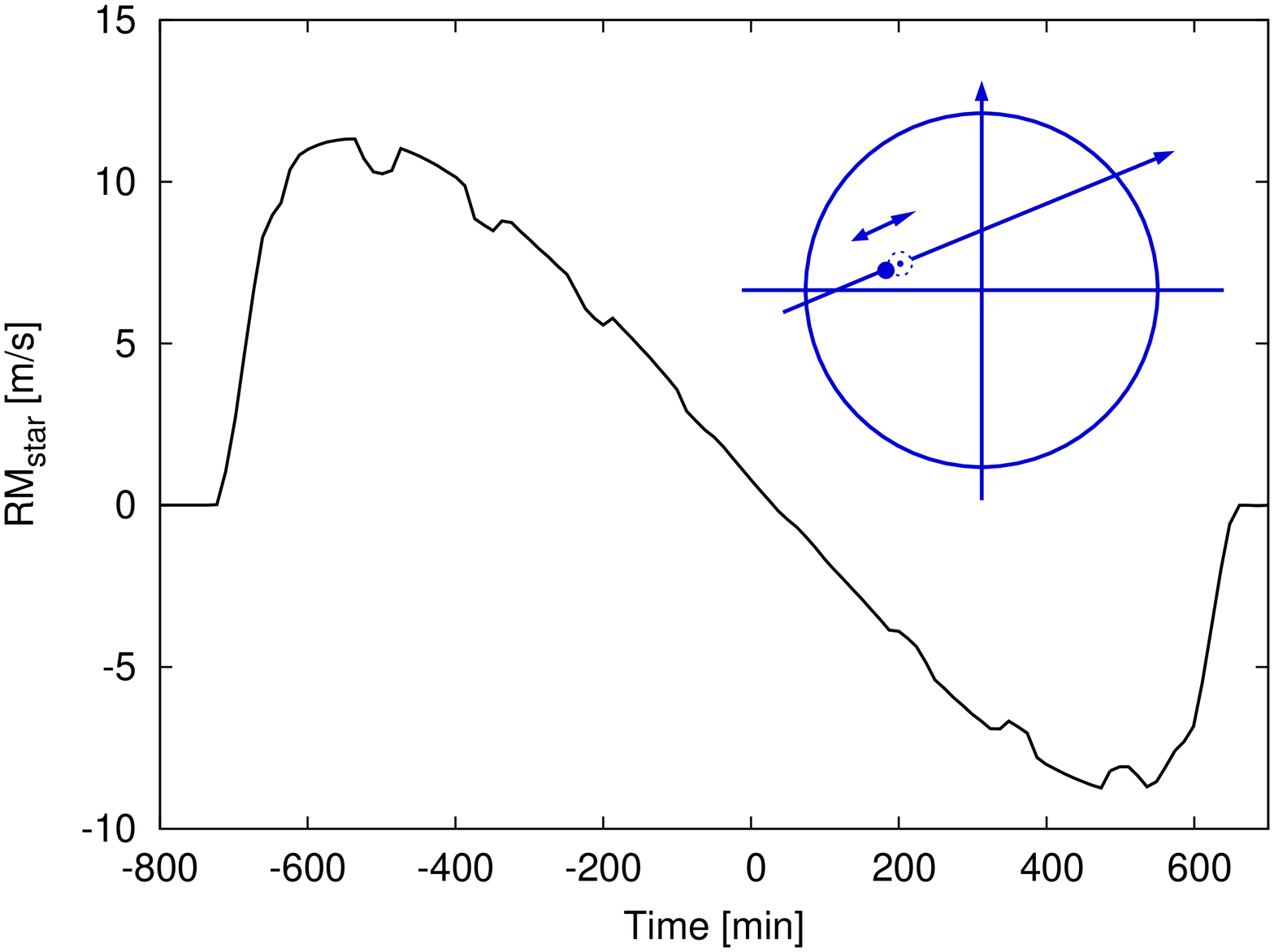}\hfill
  \includegraphics[bb=90 60 720 480,width=0.4\textwidth]{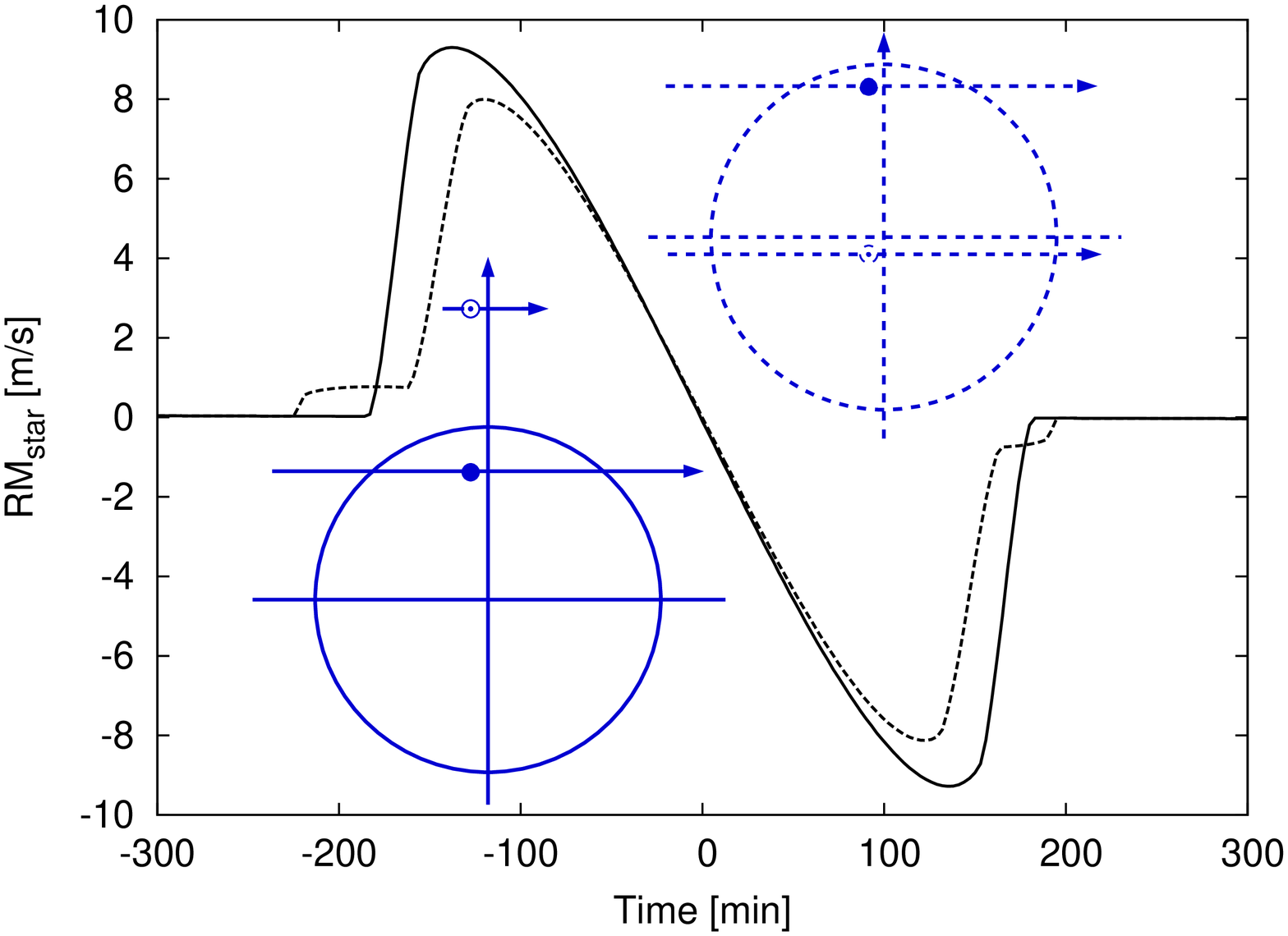}
  \vskip1cm
\caption{Four examples of the Rossiter-McLaughlin effect with different
transit geometries. 
Each panel shows the effects of the moon superimposed on the 
RM curve and the system configuration during the transit. 
The big circle with axis, the big spot and the dot in the 
small circle represent the star, the planet and the moon, 
respectively. The illustrations are to scale.}

\label{fig:2}       
\end{figure*}

We developed a new algorithm for precise calculations 
of arbitrary transiting systems with a satellite. The observed quantities
are the light curve and the Rossiter-McLaughlin curve. 
The masses, radii, orbital periods, inclinations 
and ascending nodes of the planet and the satellite are input parameters.
The orbital phase of the satellite at mid-transit is also adjustable.
The dynamical evolution of such a system is given by a three-body problem. 
At this point, we included
an approximation where the planet and the satellite  
orbit around their barycentre and this barycentre orbits 
the star with uniform velocity. To exclude the short-time 
escape of the moon, the system had to fulfil the criterion on the 
Hill sphere and must orbit at larger radius than the Roche 
limit (Szab\'o et al. 2006).

The star is parametrized by its mass and limb darkening 
(Phoenix linear limb darkening coefficients are default, Claret 2000).
Stellar radius is calculated 
from the mass using the models of the Padova isochrones 
with solar metalicity (Girardi et al. 2002).
The apparent brightness and the radial velocity of the star change
while the satellite transits. These observables are calculated by integrating over
a simulated and discretized stellar disk. Where the
planet and the satellite hide the stellar surface, pixels get zero weight.
We use a model star with a radius of 1000 pixels. The pixels are rectangular, 
and each has two values: (i) local surface brightness and (ii) local velocity 
of the surface element, calculated from the radius, the stellar spin period
and the position of the pixel.

The radii of orbits (planet and satellite) are calculated 
from the orbital periods, according to Kepler's third law.  
The velocity vectors of the planet and the moon are derived from their 
position vectors and their orbital periods. These determine the radial velocity
of the barycentre of the star via the criterion of the conservation of
momentum, leading to a simple dynamical dumbbell-model.

This algorithm is implemented in a user-friendly GUI interface\footnote{Under
Labview Environment, NI Instruments, www.ni.com/labview} with four animated
simulation windows in a shared Front Panel. They show (i) a distant view of the orbit of the
planet and the satellite; (ii) a zoom into the transit geometry; (iii) 
the surface of the star with the transiting objects and (iv) the light 
curves and RM curves. The input parameters, stability information
and applied time step are indicated in the parameter panel (main window,
Simon et al. 2009).

The user can set the time step
and can select real time or background calculations. 
It is also possible to generate a large number of of photometric 
and RM curves using system parameters randomly within a given 
interval. The output files contain a detailed 
file header with the system data and three 
columns with values of time, magnitude and radial velocity data. 



\subsection{Sample simulations}

We simulated many systems with various parameters, and identified 
different transit scenarios that lead to morphologically different light
curves and Rossiter-McLaughlin curves. This will lead to a classification
of transit geometries from an observational point of view.
First we simulated
the two most prominent examples in our Solar System: the transit of Earth with
the Moon, and the transit of the Jupiter with Ganymede. The light curve and
RM curves are presented in the upper panels of Fig. 1. We concluded that the
photometric effects of these moons are too little for a direct detection: they
are in the order of 5--10 ppm. This is an order of magnitude less than the 
photometric accuracy of the Kepler space telescope (255 
ppm $rms$ in the time series of 115 quiet stars in Short Cadence data, 
Gilliand et al. 2010). 
The RM effect due to the moon is again very small, 1 cm/s and 4 cm/s, respectively. 
These velocity deviations are beyond the current technology and also are also much 
smaller than the intrinsic stellar noise for a wide range of stellar parameters (cf. Sect. 4). 
For these configurations, the only chance for the detection is the Photocentric Transit
Timing Variation (Szab\'o et al. 2006, Simon et al. 2007).
If the moon is so close that during one transit it orbits the planet more than once,
this can lead to marked waves (wobbling)
in the RM curve (Fig. 2 lower panels) that can have an amplitude of 10--100 cm/s.

The lower panels of Fig. 1 show systems where the detection is more promising. We designed systems with a little star, and very large, Earth-sized moons
of transiting Saturn-like planets. The data of the systems is summarized in
Table 1. Here the effects of the moon can be as large as 730 ppm in photometry,
or 80 cm/s in radial velocity. Both values are promising, so the conclusion is
that if such systems exist, they could be discovered with the present techniques. 
In these particular examples the
moons orbit very close to the planet, mutual planet-satellite eclipses 
occur, but the presence of mutual eclipses is not a necessary criterion for a successful
discovery. The most important parameter is the size of the moon, that must be in
the order of the size of the Earth. This is the size that can cause directly
observable effects, almost regardless of the orbital period of the satellite
itself.

\begin{table}
\caption{Transit classification and proposed methods for detecting the moon.
LC: direct detection in the light curves, TTV$_{\rm b}$ and TTV$_{\rm p}$: 
barycentric and photocentric TTV, TDV: Transit Duration Variation, RM: the Rossiter-McLaughlin
effect, RM(TV): Transit Timing Variation of the RM effect.}
\label{tab:3}
\begin{tabular}{llll}
\hline\noalign{\smallskip}
 & Static  & Slow & Rapid \\
  & moon          & moon & moon\\
 \hline
 Full & LC, TTV$_{\rm b}$, & LC, TTV$_{\rm b}$, & LC
 (wobbling),\\
 Transit & TTV$_{\rm p}$, RM &  TTV$_{\rm p}$, TDV& RM (wobbling)\\
         &                   & RM\\
 \hline
 Tangent \\
 planet & LC, RM & LC, RM & LC, RM\\
 \hline
 Planet & TTV$_{\rm b}$, & TTV$_{\rm b}$, TDV&  TTV$_{\rm b}$, TDV,\\
 only & RM(TV) & RM & RM\\ 
\hline\noalign{\bigskip}
\end{tabular}
\end{table}

\subsection{Classification of transit scenarios}

By now, there are several methods that offer the detection
of exomoons, while they are not equivalently effective for different transit
configurations. A possible classification scheme for transits is proposed here from the point
of view of the applicable methods.

We have run many transits with the purpose of mapping the entire
parameter space. These lead to different morphology of light curve and RM curve, therefore
different methods are required to detect the moon itself in the measurements.

There are three distinctly different geometrical configurations which we call as the ``static'' moon, the ``slow'' moon and the ``rapid'' moon. A ``static'' moon means that it has a very long orbital period, comparable to that of the planet itself. In a such configuration, the relative position of the moon to the planet does not change significantly during one transit. Thus, one transit of the planet and one transit of the moon, superimposed to each other, are observed.
The shape of the light curves and RM curves are the same for
the planet and the moon, but time-lagged because of the geometry (upper left panel in Fig. 2). 
In extreme cases, e.g. when the semi-major axis of the moon is greater than the stellar diameter, the planet may complete the entire transit before the moon contacts the stellar disk, in which case we observe totally distinct ``planet"
and ``moon" transits -- the latter with a much smaller depth.

In the case of a ``slow'' moon, a difference can be detected
between the duration of the transit of the moon and that of
the planet, due to the slowly changing relative position during the event.
This is favourable because the orbital period can also be estimated
from one single transit observation (cf. Sect. 4).

In the two cases described above there are some scenarios when only the planet or the moon transits completely, resulting in a short grazing eclipse of the stellar disk by the planet or the moon (``tangent planet'' configuration, see in Fig. 2, top right). There is another setup when the moon does not transit at all, but causes TTV and TDV in the transit of the planet. The lower right panel in Fig. 2 shows two such transits with the moon on the opposite sides of the planet.

In the third case the satellite orbits rapidly, which produces a characteristic wobbling in the light curves and the RM curves. The great variety of such extreme geometries is interesting but the large diversity of possible scenarios does not enable drawing a consistent picture (see Fig. 2, bottom left, for an example).

These configurations prominently differ and give different observable effects.
In Table 2 we summarise all the cases, together with the methods that can be applied to detect the satellite.
In the following we restrict the discussion to the slow and static moons.


\section{Inversion of the Rossiter-McLaughlin effect}

The task of analysing real observations is to recognize
the presence of an exomoon if it exists, and to estimate 
its parameters and parameter errors. 
Here we present an elaborated analysis of parameter reconstruction,
correlations and degeneracies from noisy simulated observations 
utilising bounded error analysis.
The transiting model system consists of a Uranus-sized planet
with one large, 
Earth-sized moon around a 0.8 M$_\odot$ mass star (see data of 
Simulation 5). The RM curve has been 
equidistantly sampled in 3 minute stepsize
and then noisified with
various amount (20, 50, 100 cm/s) of uniform noise. 
While the RM effect of the moon itself had
1 m/s amplitude, these refer to approximate 
signal (of the moon) - to -noise ratios of 5, 2 
and 1, respectively.

It is worth noting that we do not claim that the RM effect is efficient enough for discovering exomoons.
Our aim here is to understand which parameters can be extracted from RM observations,
regardless of to what extent they can be constrained from transit photometry.

A joint analysis of the RM effect and transit photometry is beyond the scope of the present paper
and will be discussed in a forthcoming publication.

\begin{figure}
\includegraphics[width=8cm]{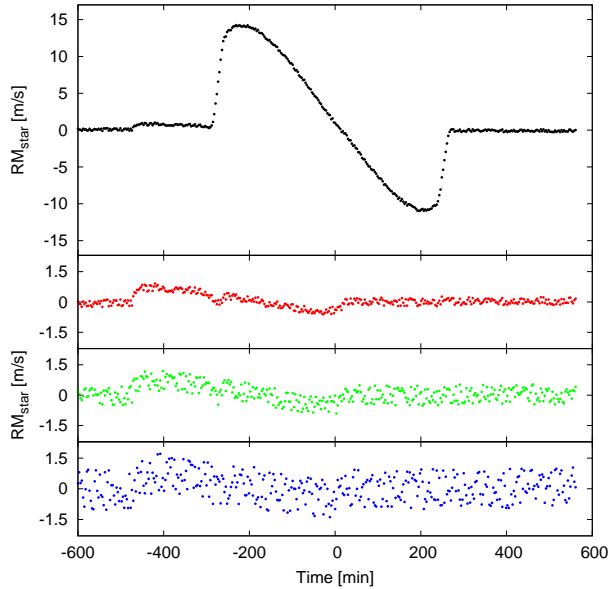}
\caption{A simulated observation (top panel) and the residuals after fitting a
single planet solution (bottom panels). The latter plots correspond to the 
S/N=5, 2 and 1 levels.}
\end{figure}


\subsection{Parameter reconstruction}

Let $P$ and $M$ denote the real parameter vectors of the
planet and the moon, respectively, which are to be estimated
from the observations.
For this we need to locate the parameter vectors of the planet, $\tilde P_i$ 
and the moon, $\tilde M_i$ that tune
the simulator into a good agreement with the observed data.
Let $v_{r,sim}$ mean the template RM curves (they are not noisy), and
$v_{r,obs}$ be the observed data (simulated observations), which 
are noisy. The best-fit parameters of the planet, $\mathcal P$
have been determined by minimising the $rms$ scatter 
of the residuals between observations
and templates containing a single planet only:
\begin{equation}
\mathcal P = \arg \min_{\tilde P_i} \left[ \sum \left( v_{r,obs} - v_{r,sim,
\tilde{P_i}} \right)^2 \right].
\end{equation}

\begin{figure*}
	\includegraphics[bb=162 51 1935 755,width=0.45\textwidth,angle=0]{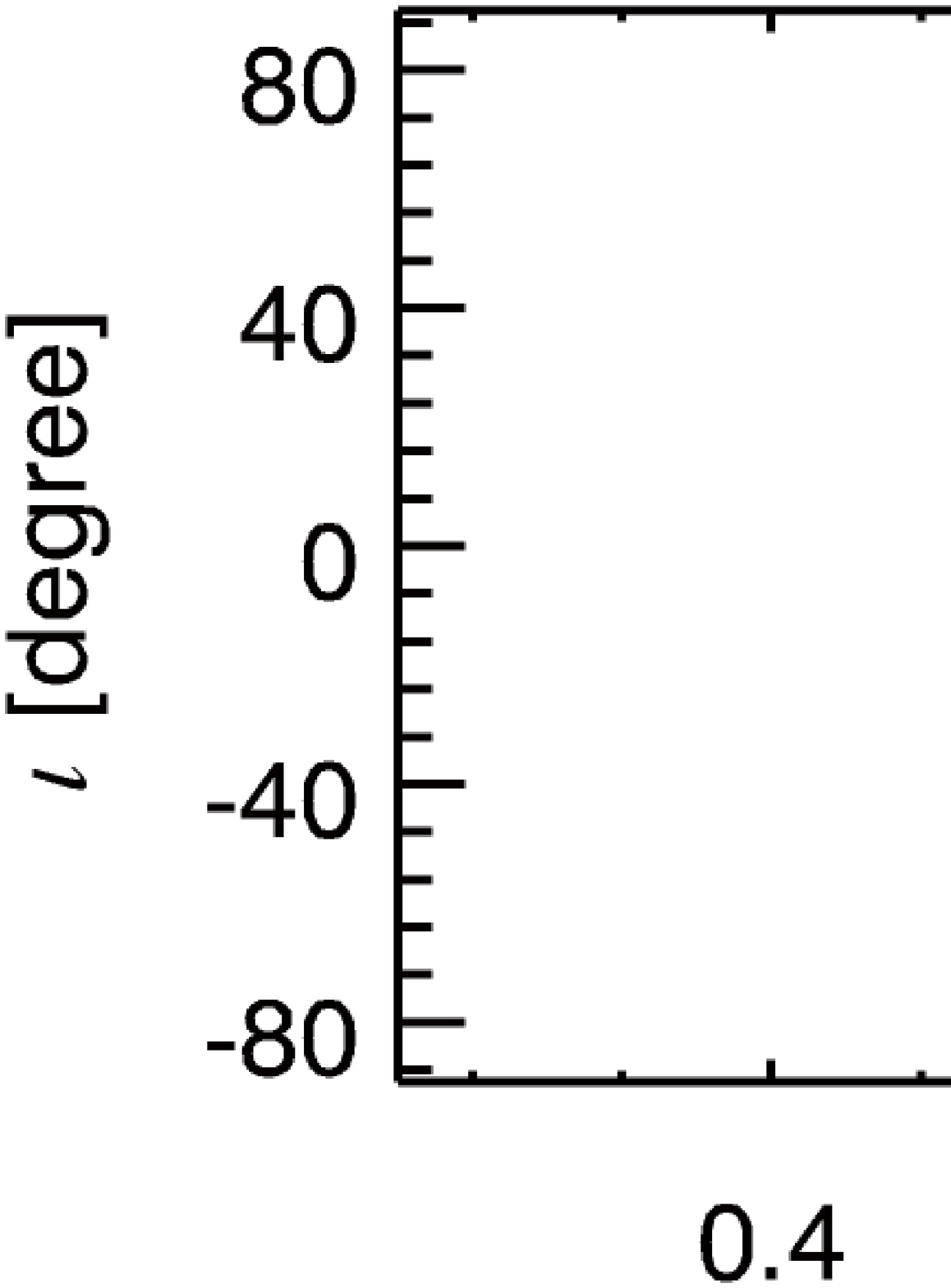}\hskip1cm
	\includegraphics[bb=162 51 1935 755,width=0.45\textwidth,angle=0]{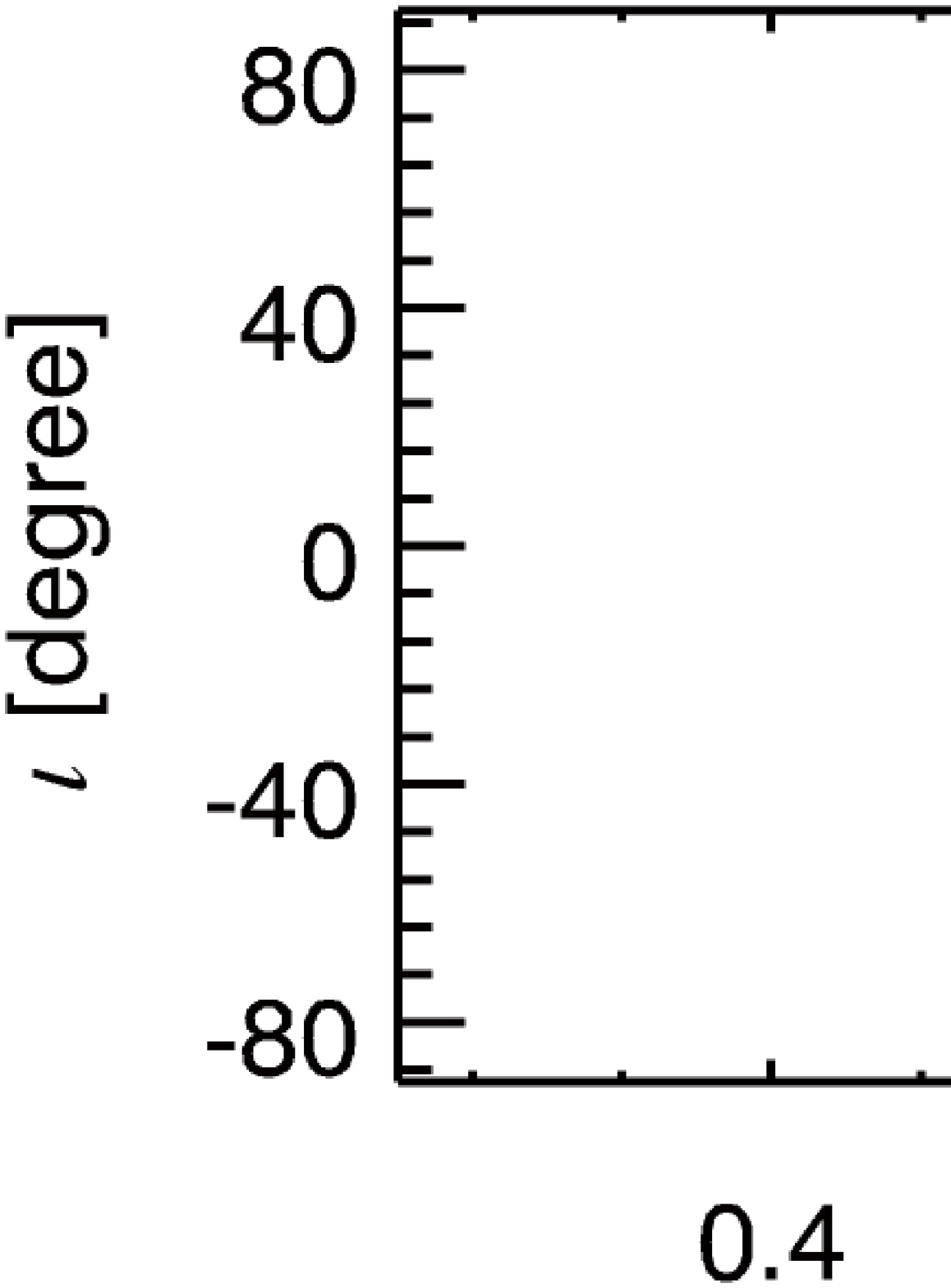}
\vskip0.7cm
	\includegraphics[bb=162 51 1935 755,width=0.45\textwidth,angle=0]{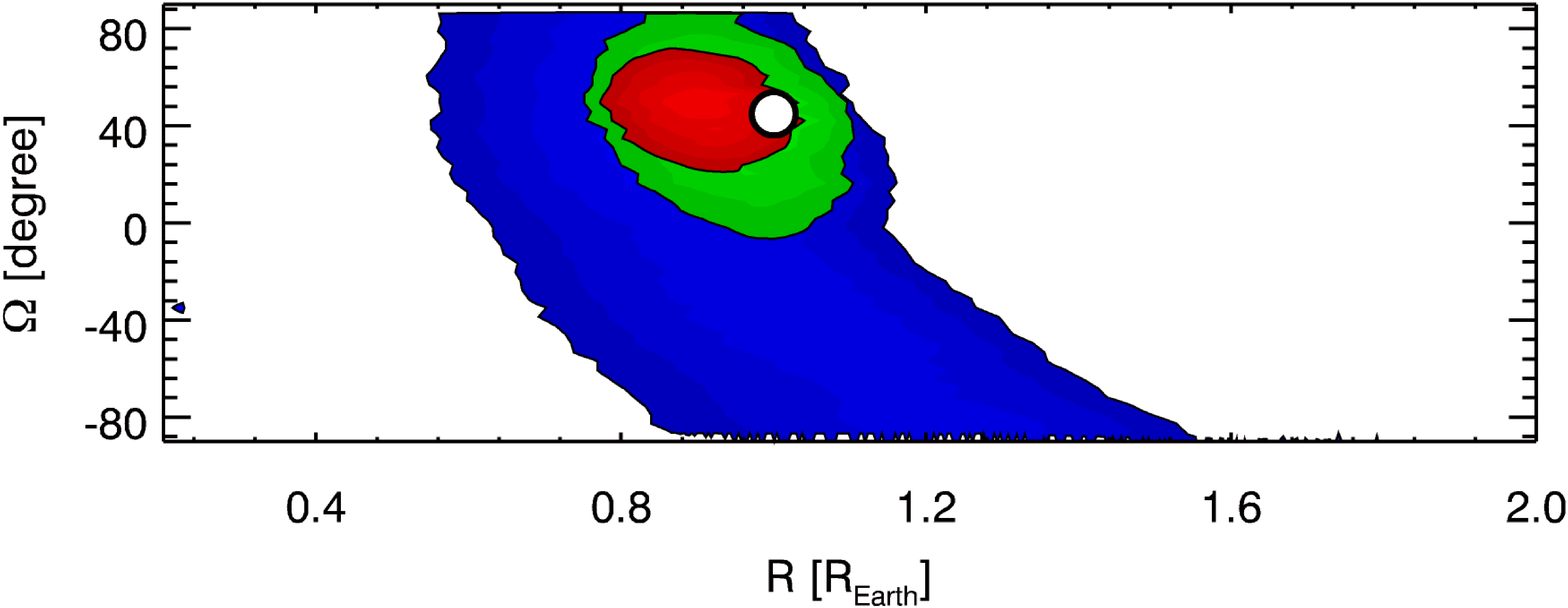}\hskip1cm
	\includegraphics[bb=162 51 1935 755,width=0.45\textwidth,angle=0]{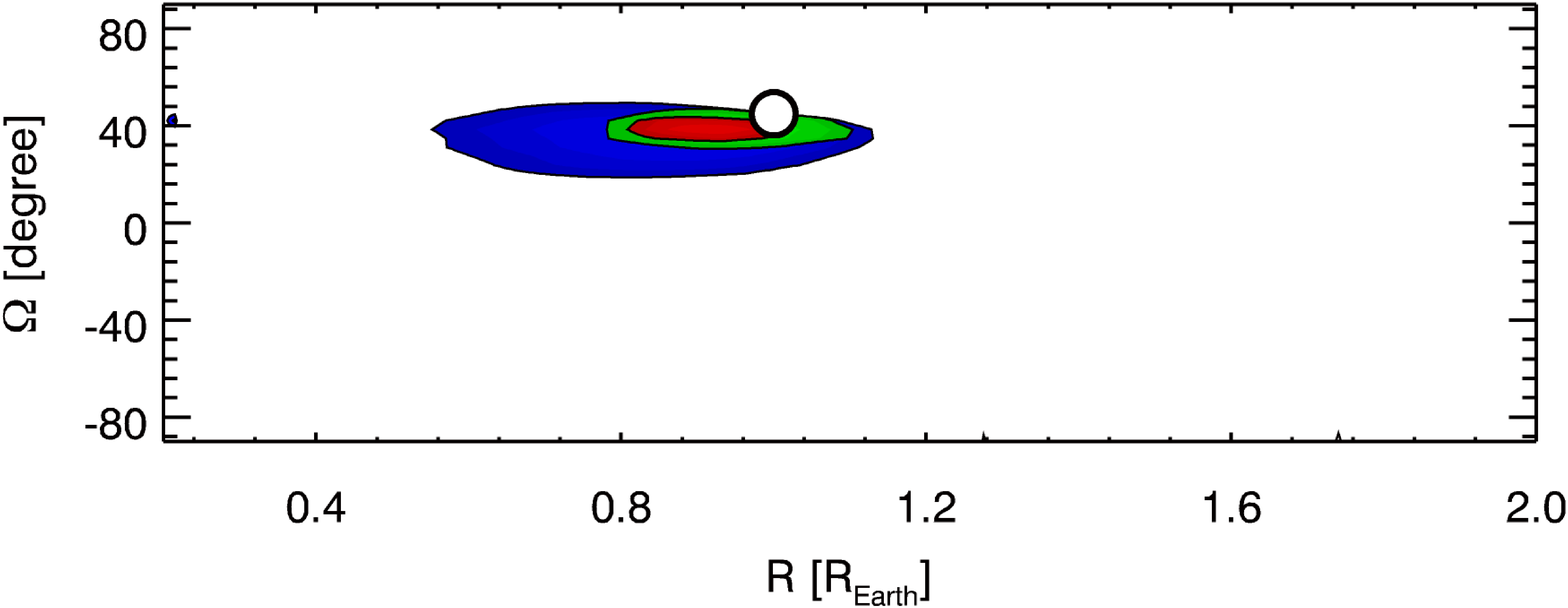}
\vskip0.7cm
	\includegraphics[bb=162 51 1935 755,width=0.45\textwidth,angle=0]{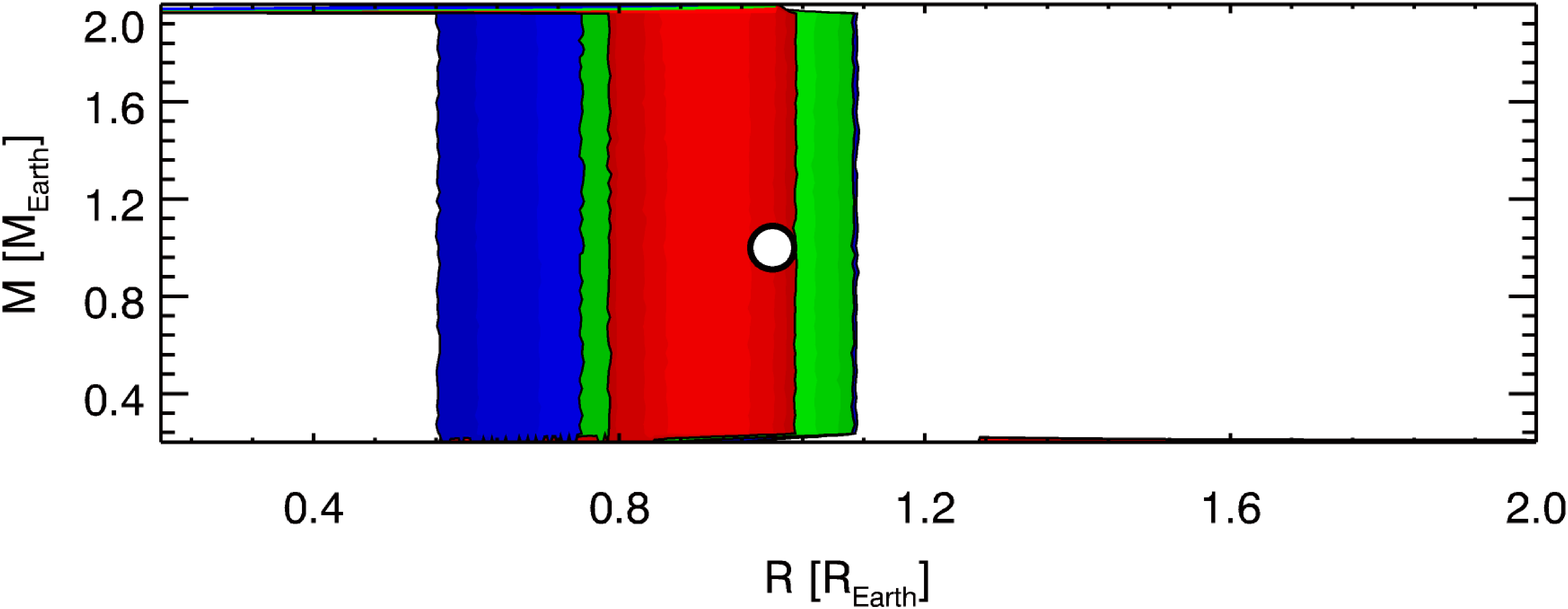}\hskip1cm
	\includegraphics[bb=162 51 1935 755,width=0.45\textwidth,angle=0]{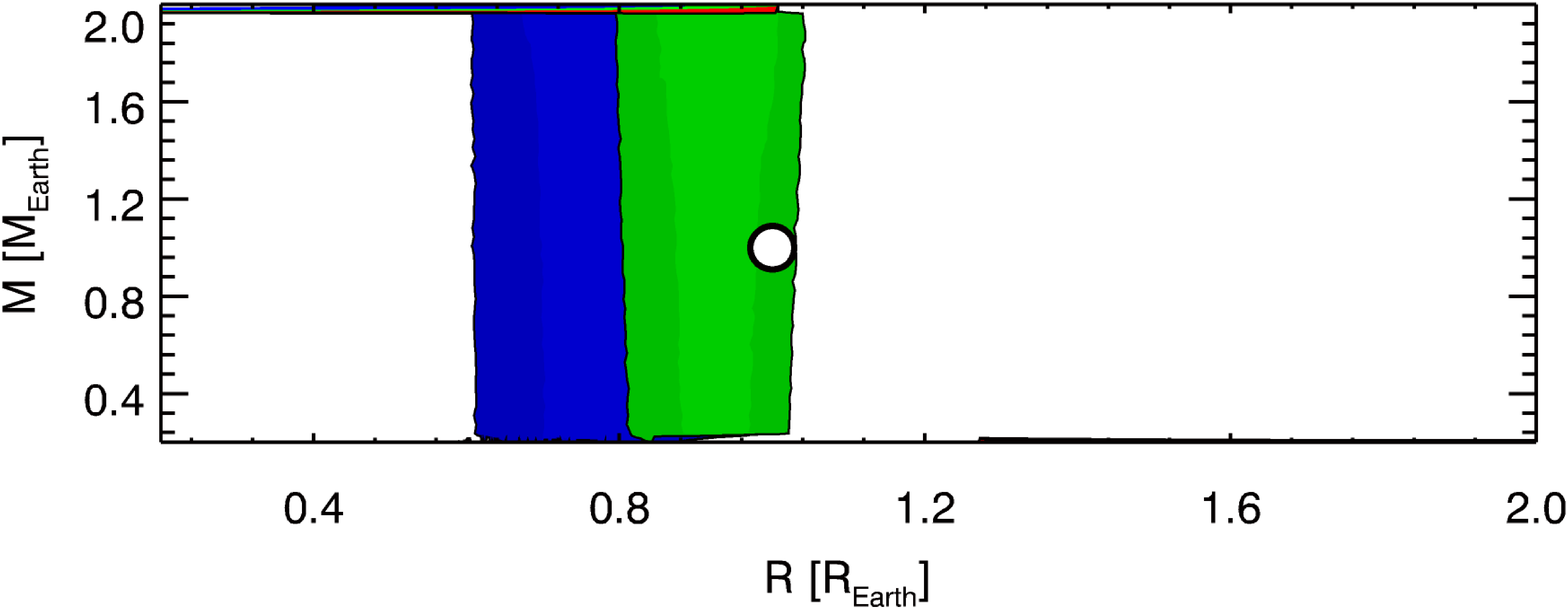}
\vskip0.7cm
	\includegraphics[bb=162 51 1935 755,width=0.45\textwidth,angle=0]{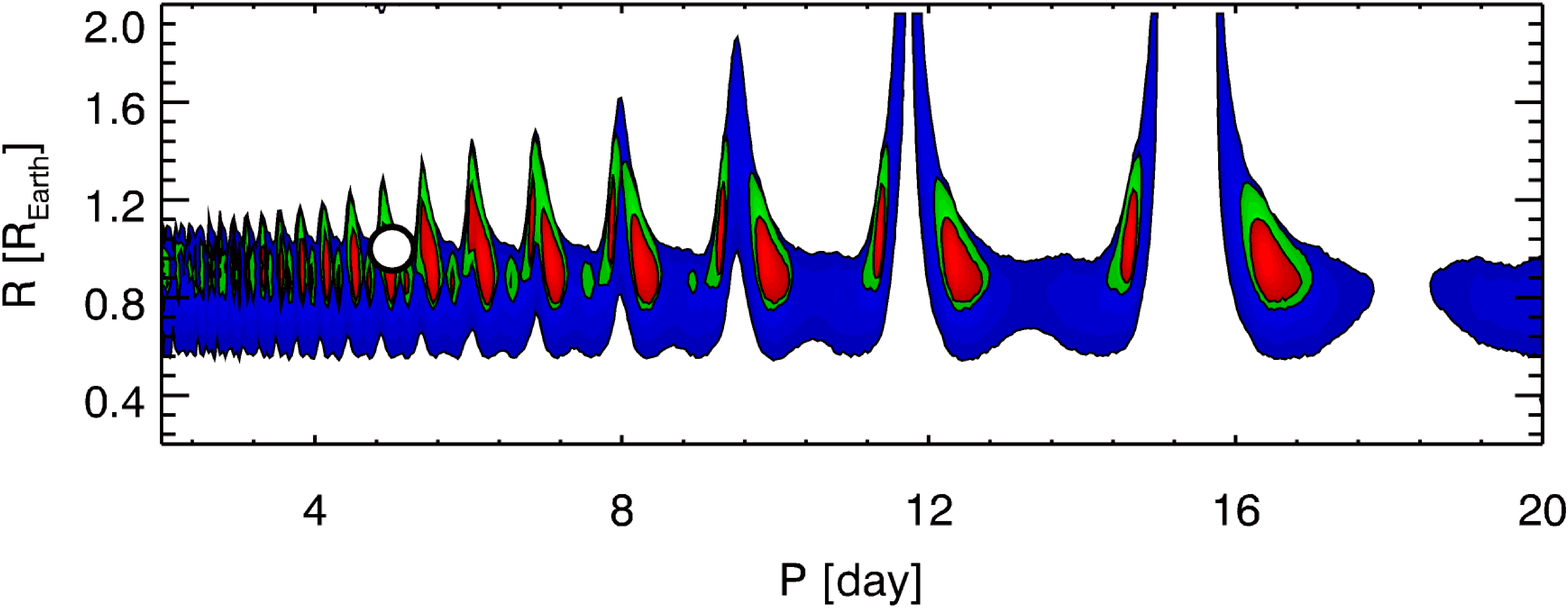}\hskip1cm
	\includegraphics[bb=162 51 1935 755,width=0.45\textwidth,angle=0]{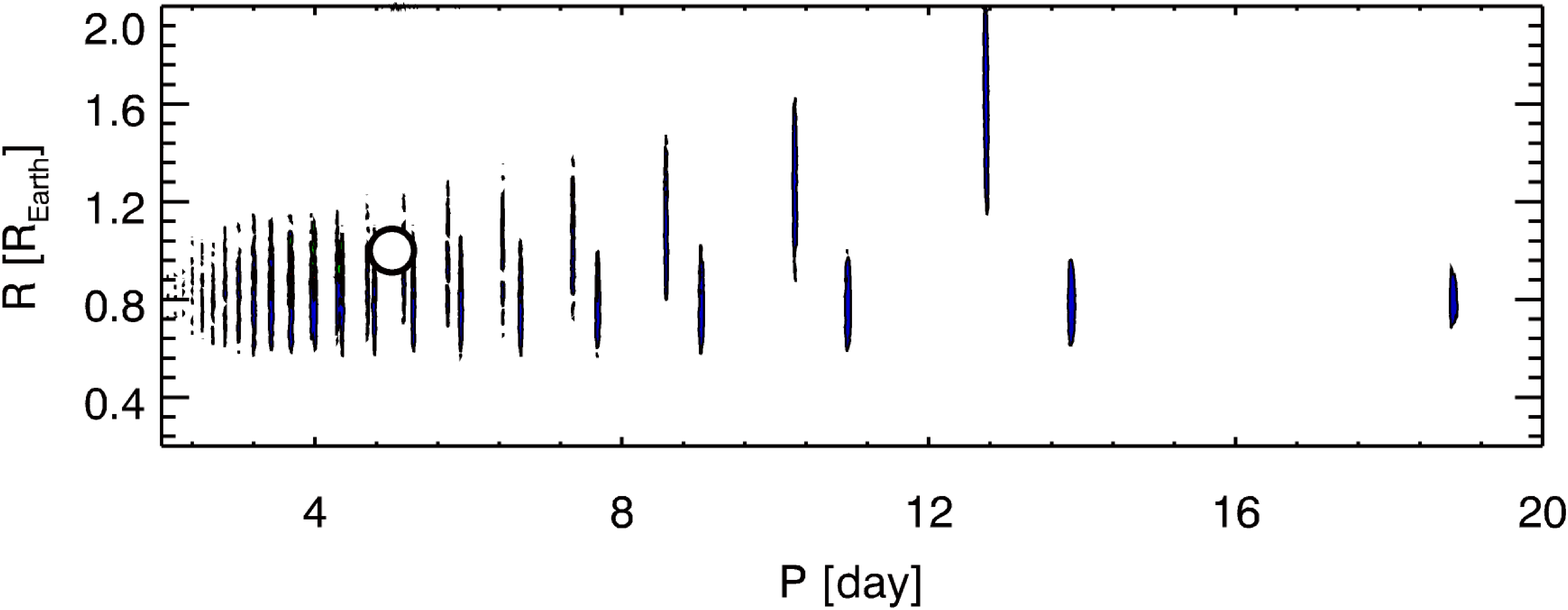}
\vskip0.7cm
	\includegraphics[bb=162 51 1935 755,width=0.45\textwidth,angle=0]{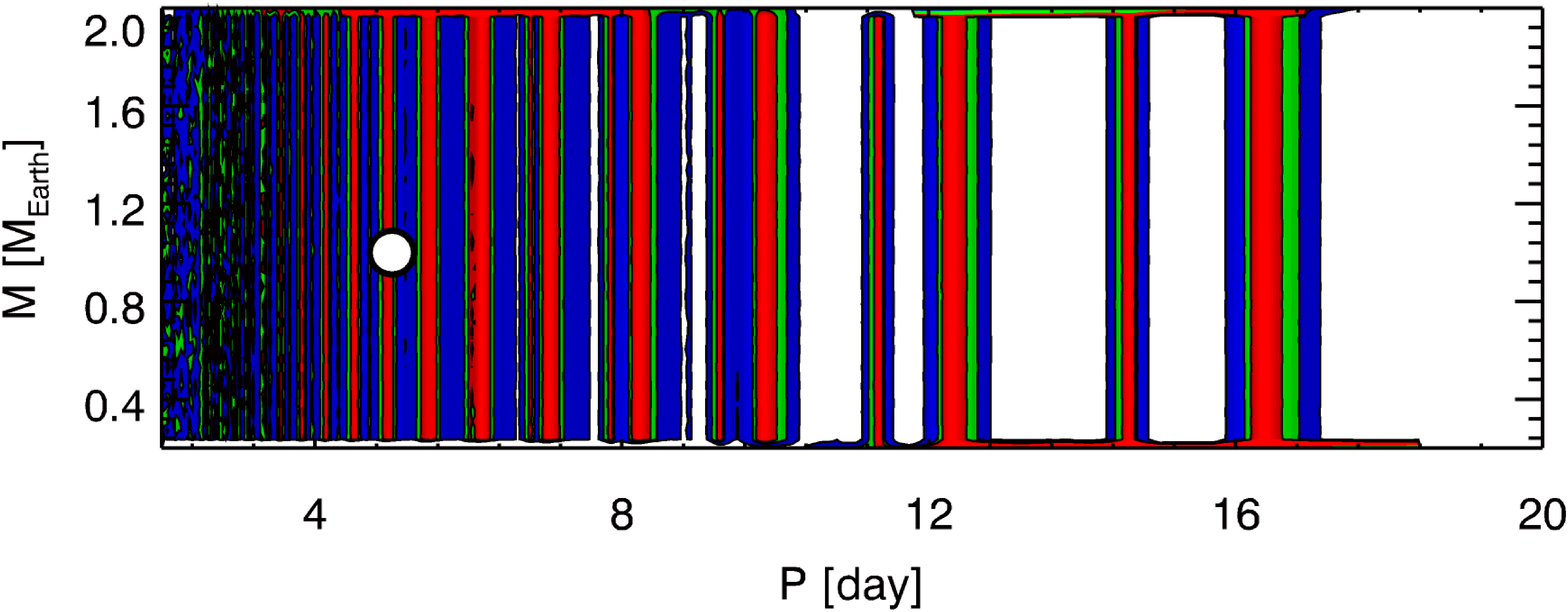}\hskip1cm
	\includegraphics[bb=162 51 1935 755,width=0.45\textwidth,angle=0]{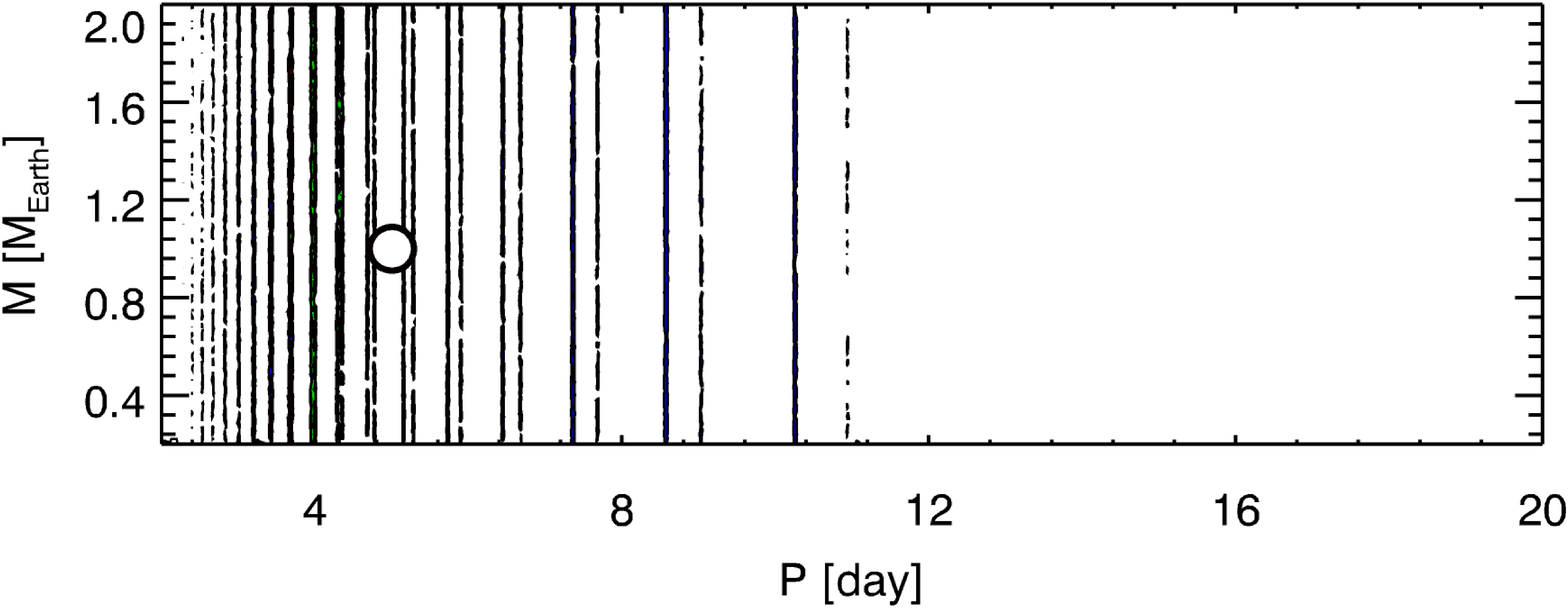}
\caption{Joint confidence intervals of the moon parameters around the exact solution. The differences between the left and the right columns demonstrate how the parameter reconstruction improves when using moon transit photometry, i.e. the central transit time of the moon is known (plots in the right columns). Different colours show fittings with different S/N ratios (red: S/N=5, green: S/N=2, blue: S/N=1), while the large open circles represent the input parameters of the simulated observations. See Table 1 for the meaning of the parameters.}
\end{figure*}

\begin{figure*}
	\includegraphics[bb=162 51 1935 755,width=0.45\textwidth,angle=0]{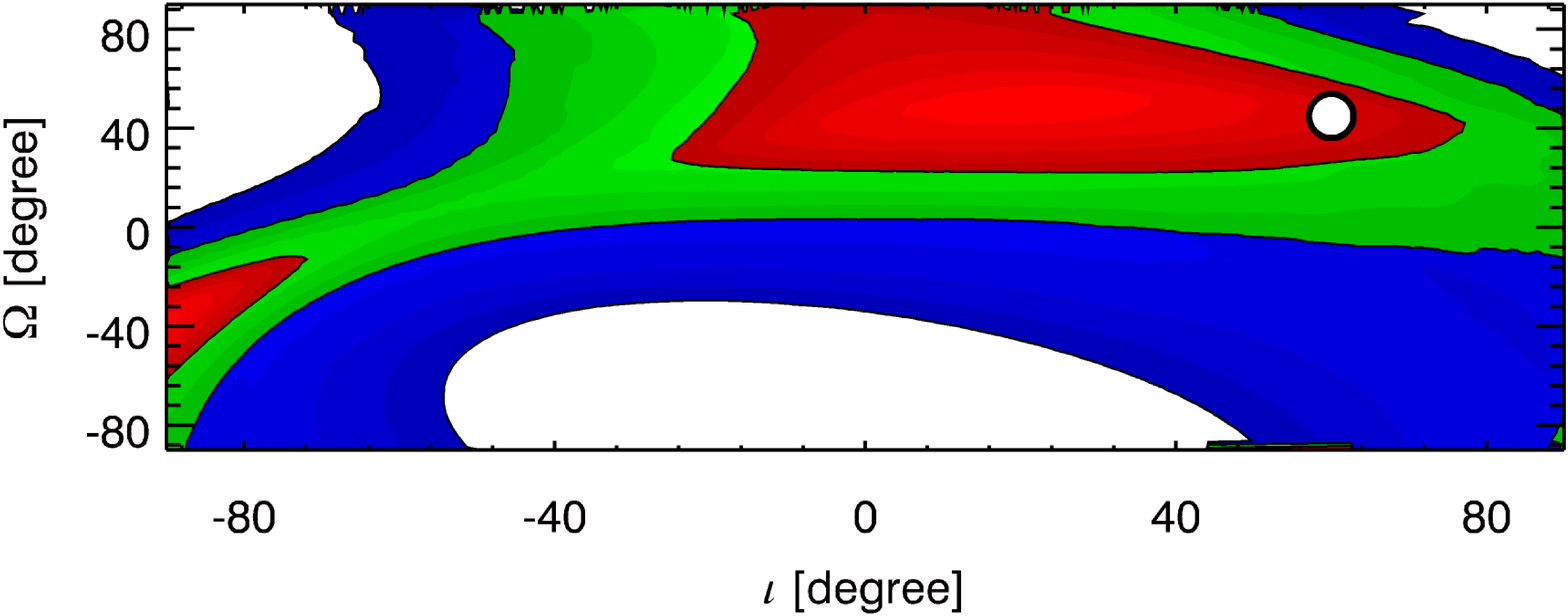}\hskip1cm
	\includegraphics[bb=162 51 1935 755,width=0.45\textwidth,angle=0]{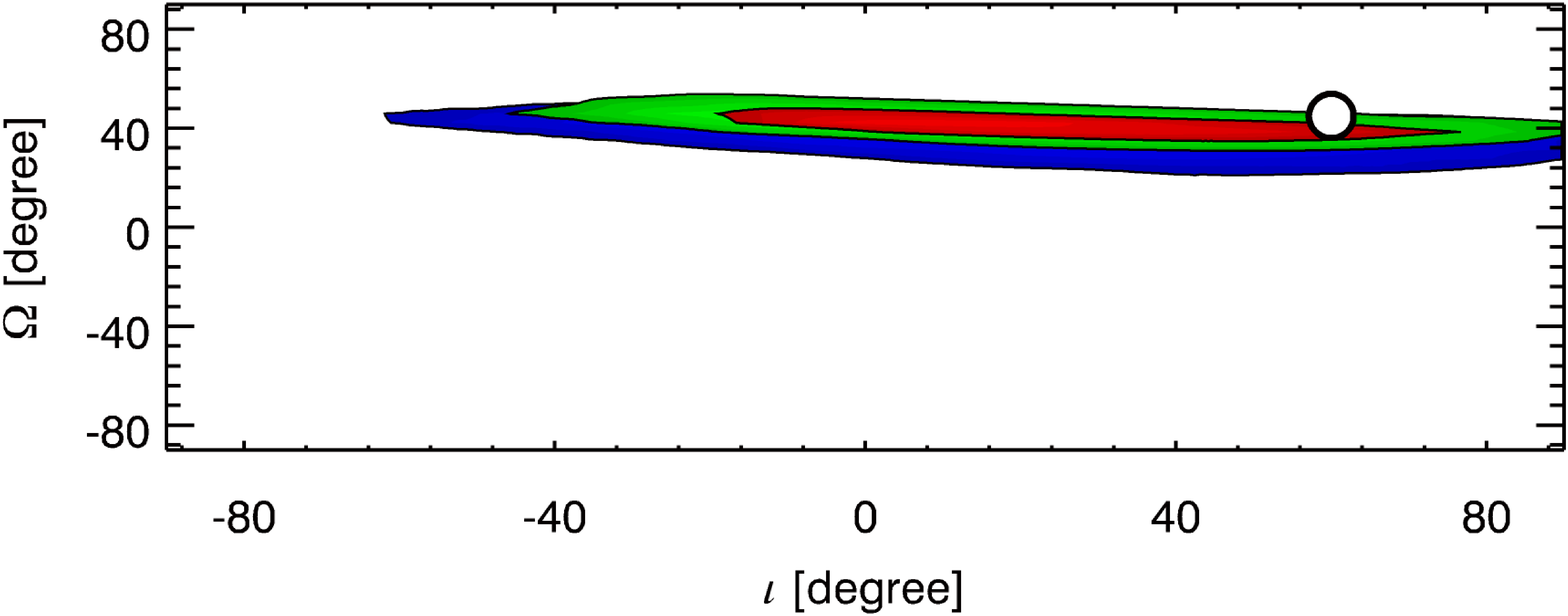}
\vskip0.7cm
	\includegraphics[bb=162 51 1935 755,width=0.45\textwidth,angle=0]{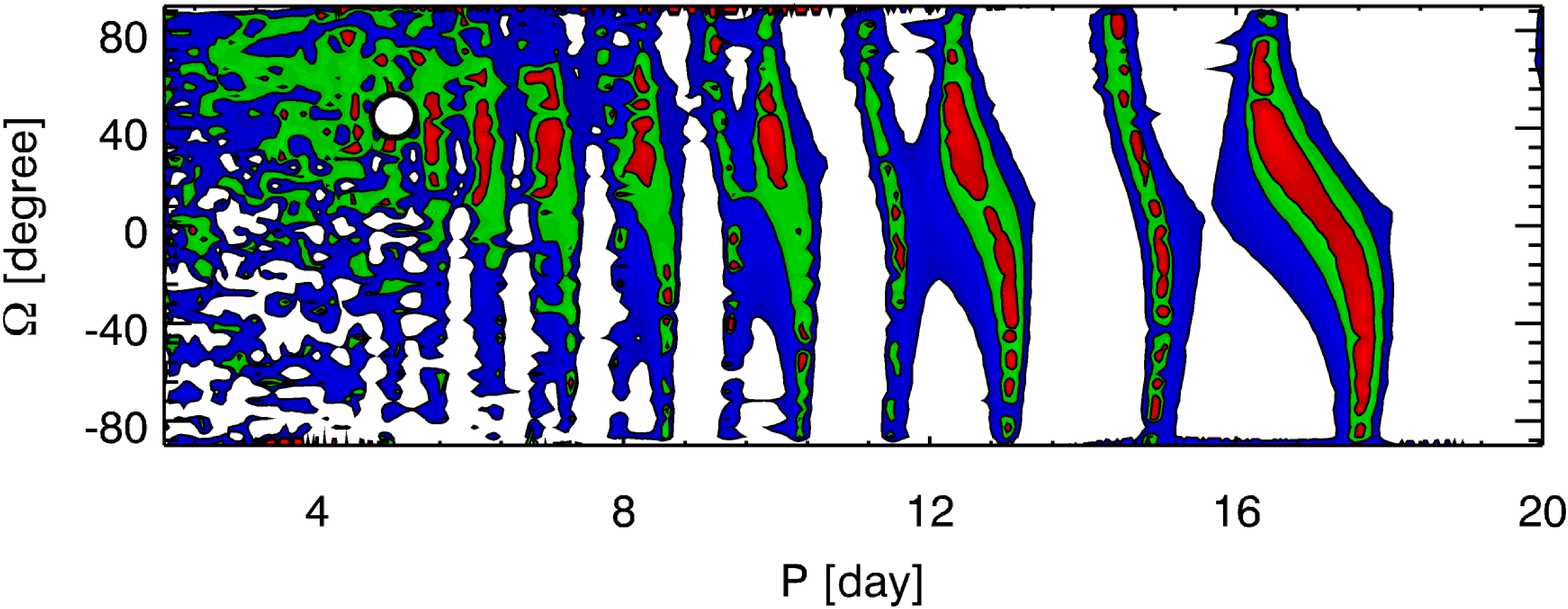}\hskip1cm
	\includegraphics[bb=162 51 1935 755,width=0.45\textwidth,angle=0]{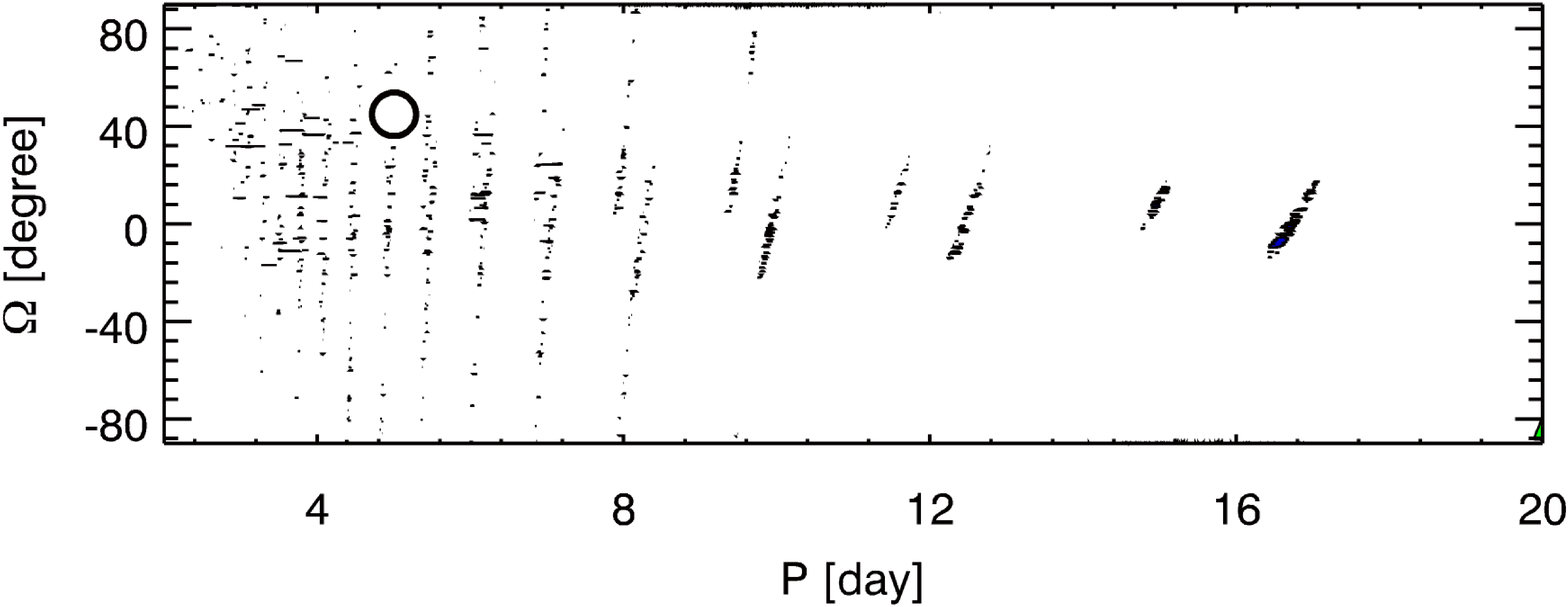}
\vskip0.7cm
	\includegraphics[bb=162 51 1935 755,width=0.45\textwidth,angle=0]{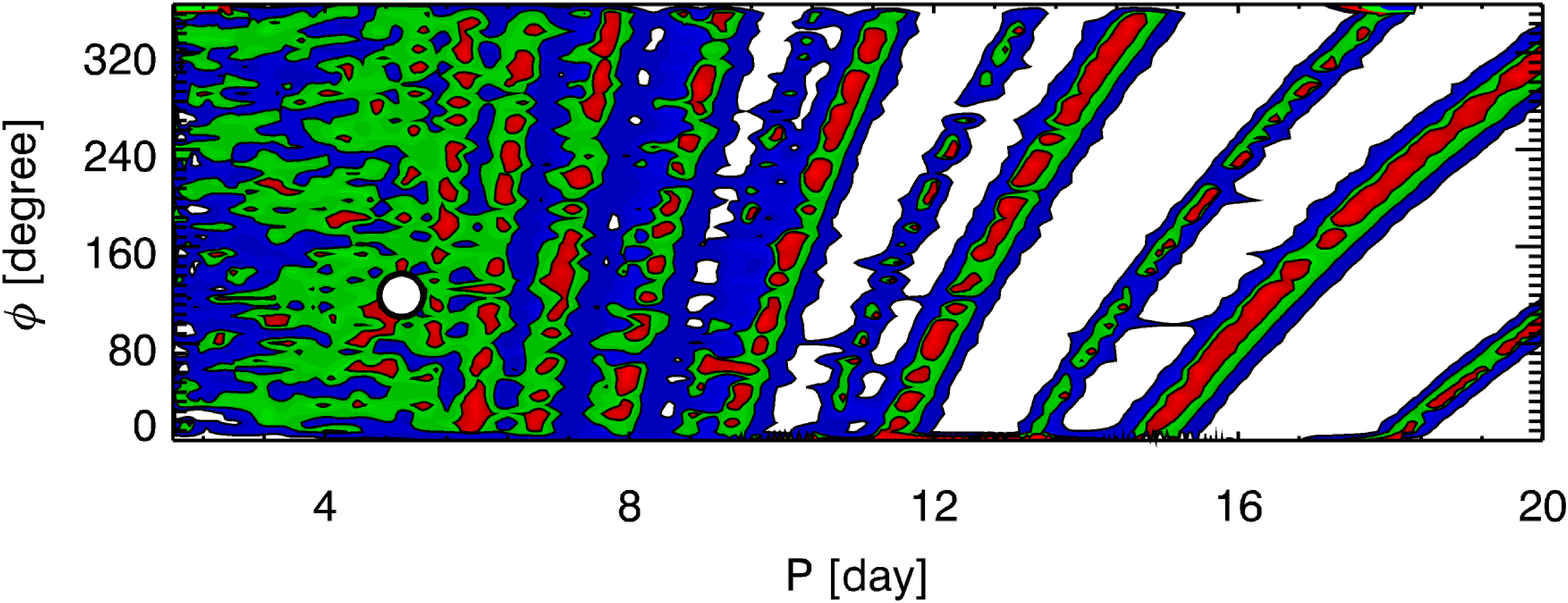}\hskip1cm
	\includegraphics[bb=162 51 1935 755,width=0.45\textwidth,angle=0]{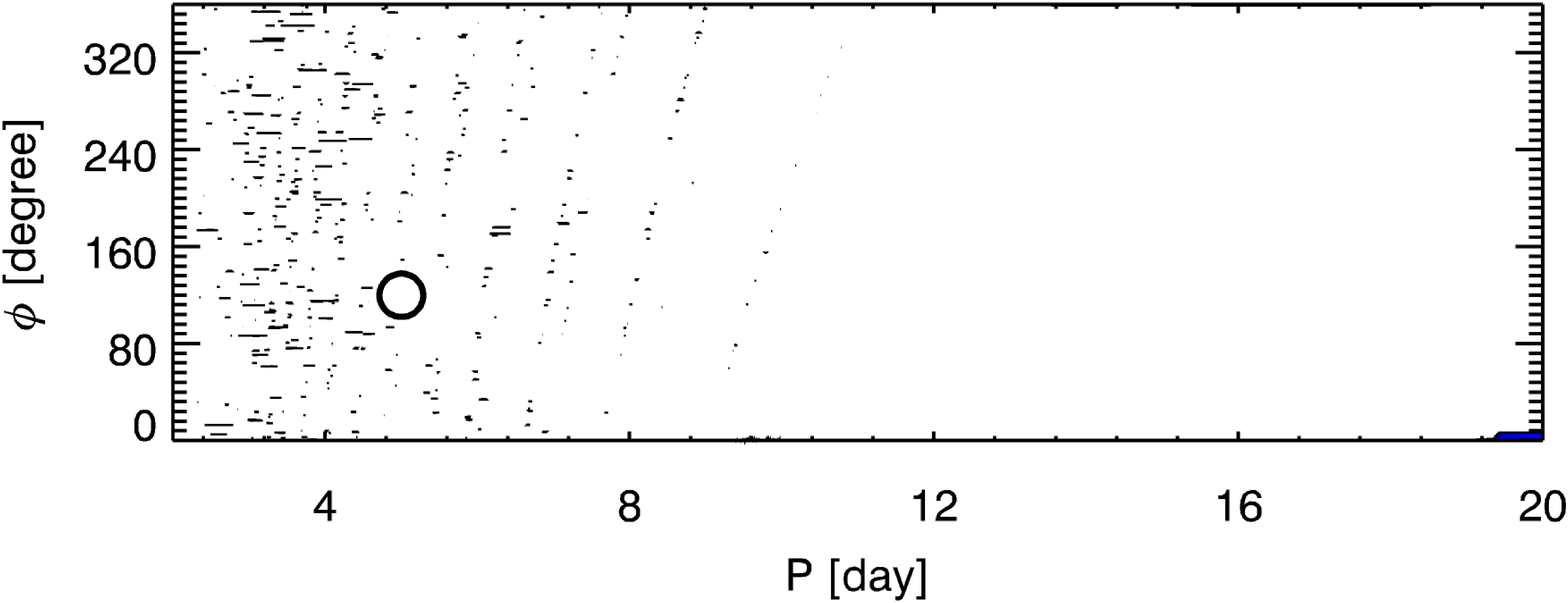}
\vskip0.7cm
	\includegraphics[bb=162 51 1935 755,width=0.45\textwidth,angle=0]{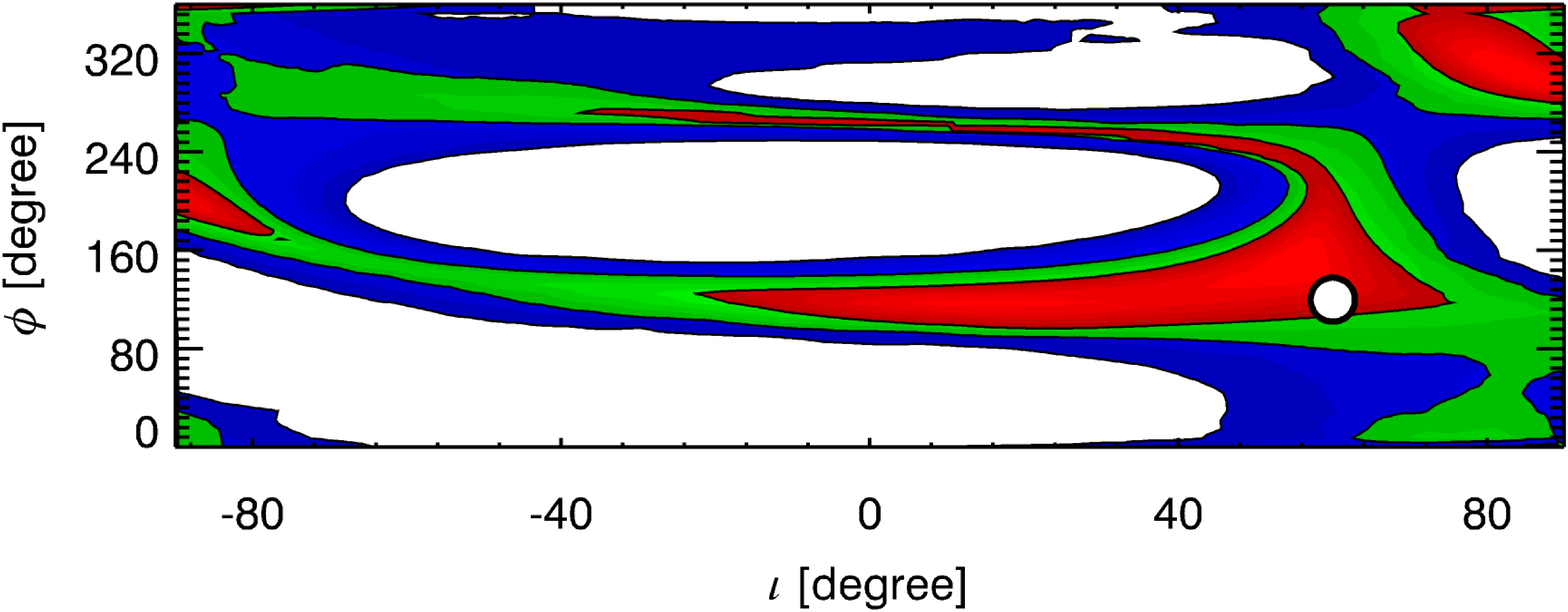}\hskip1cm
	\includegraphics[bb=162 51 1935 755,width=0.45\textwidth,angle=0]{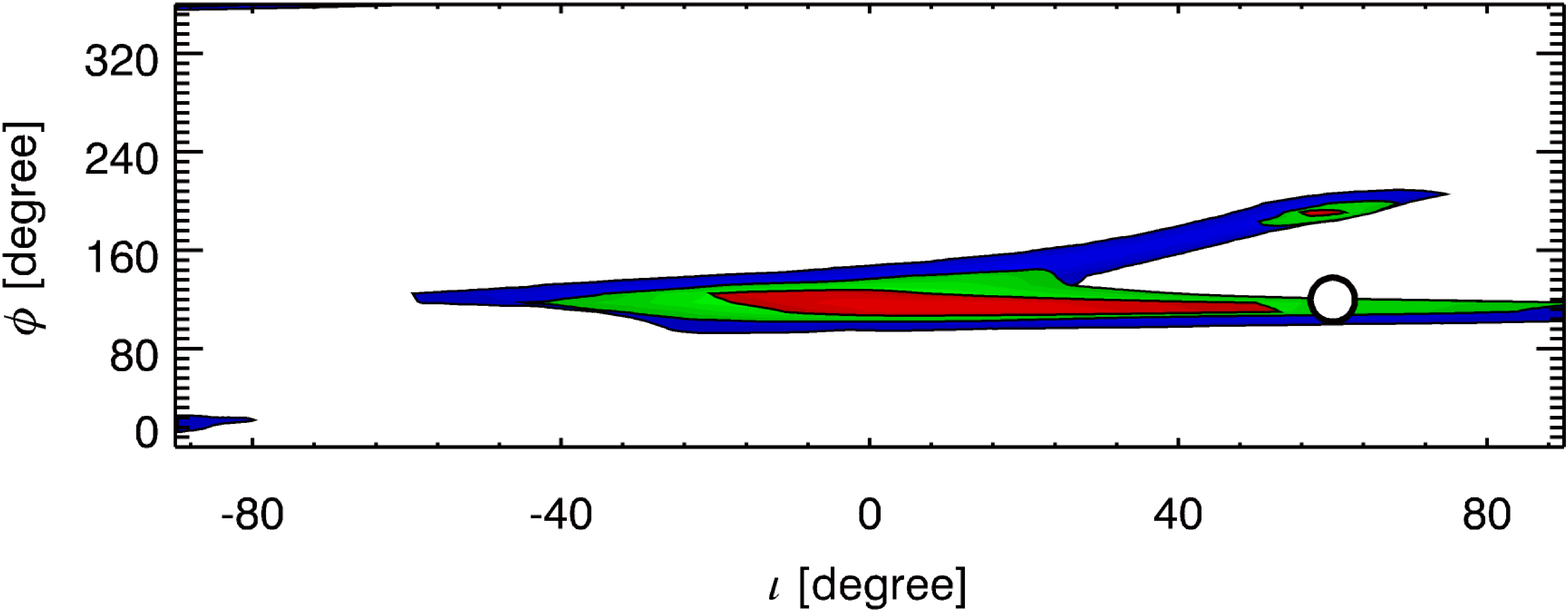}
\vskip0.7cm
	\includegraphics[bb=162 51 1935 755,width=0.45\textwidth,angle=0]{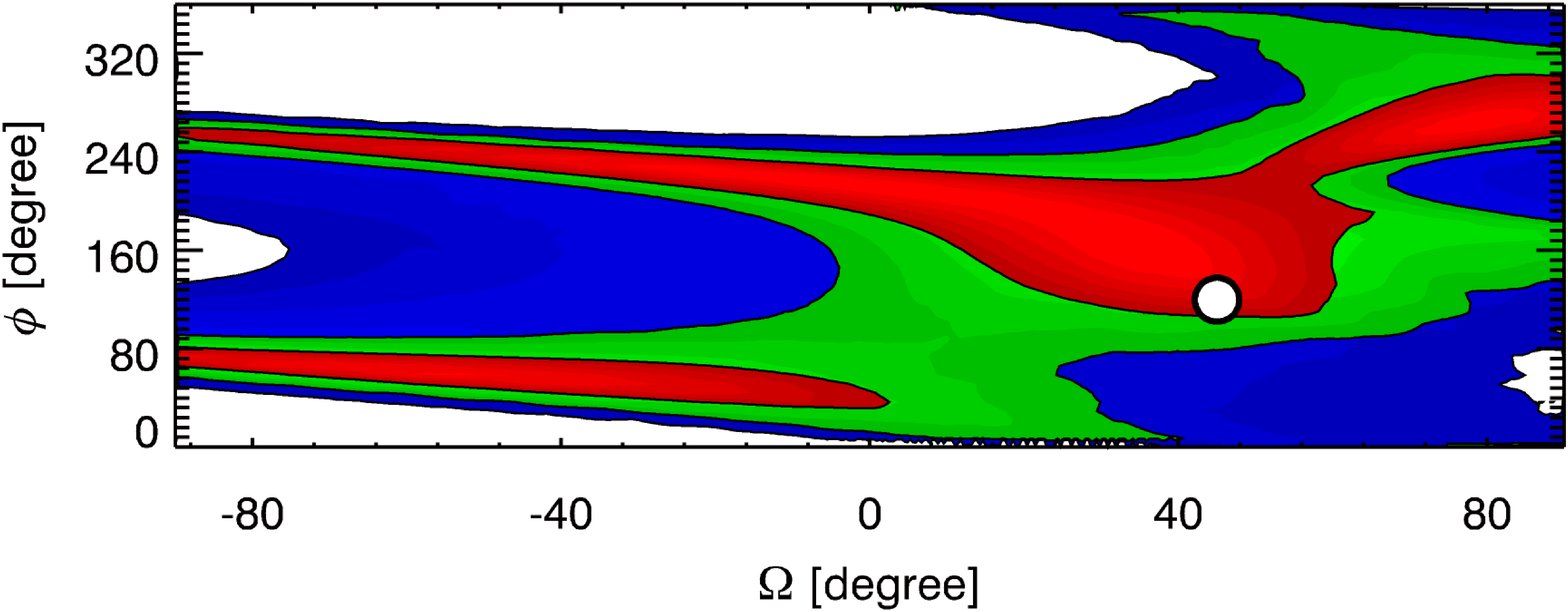}\hskip1cm
	\includegraphics[bb=162 51 1935 755,width=0.45\textwidth,angle=0]{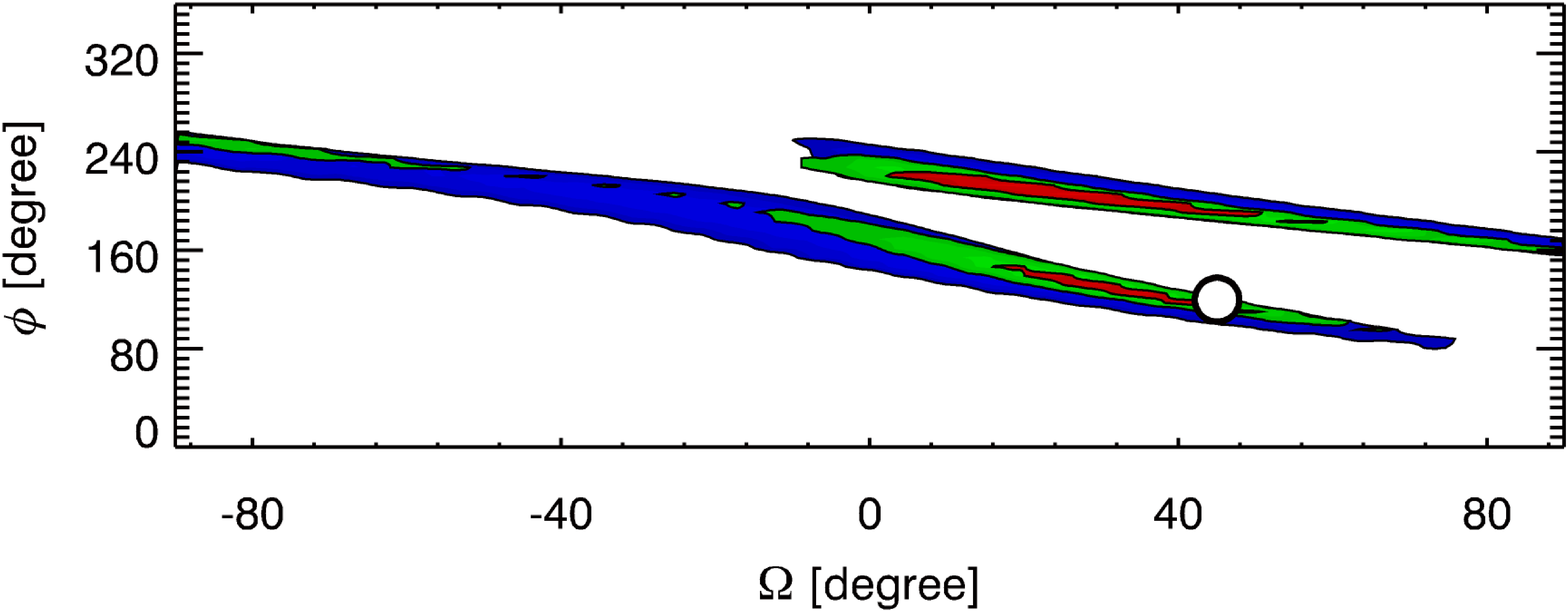}
\caption{The continuation of Fig. 4. Here  $\phi$ means the orbital phase of the moon around the planet.}
\label{fig:4}       
\end{figure*}

We determine $\mathcal M$, the parameter vector of the moon in the second step.
The residuals between the observations and the planet template contain the
signal from the moon and noise. We fit this residual with tuning 
the parameters of the moon:

\begin{equation}
\mathcal M = \arg \min_{\tilde M_i}  \left[ \sum \left(  
 v_{r,obs} - v_{r,sim,\mathcal P}  - 
v_{r,sim,\mathcal P, \tilde{M_i}}
\right)^2 \right].
\end{equation}

In this formulation 
we decoupled the parameter estimation of the planet and that of the moon. Our
tests have proven that this can be done. This is 
because the RM signal of the moon is
only a slight perturbation in the RM pattern of the planet. Thus, the planet 
parameters can be reconstructed accurately enough that they do not influence
the selection of the appropriate moon model in the second step.

\subsection{Error analysis}

After determining the parameters of the
star and the planet, there remain 6 independent parameters
of the moon.
The confidence region in this 6-dimensional 
hyperspace has been mapped in 2-dimensional plane
sections that the 15 possible parameter-parameter 
pairs span. $A_{x1,x2}$ is the joint confidence region 
(region of acceptance) of $x_1$ and $x_2$ in the 
$X=x_1,x_2,...$ hyperspace if
\begin{equation}
\int_{A_{x1,x2}} P\big( $\rm$obs \ \big| X ) \ \ {\rm d}X = C,
\end{equation}
where $obs$ means the observed data set with errors, and 
$C$ is the level of confidence. The smallest confidence interval
at a given $C$ confidence is conjured by a certain likelihood value
that can be determined experimentally.
The  $P()$ probabilities are calculated from the reduced standard 
deviations of the fits, using $\chi^2$ statistics. 
With this definition of the  confidence 
interval the correlations can be recognized
easily, because correlated parameters depend a lot on each other 
in the region of acceptance. A parameter that can be well reconstructed
must have small confidence interval in all sections, and must not be
strongly correlated to other parameters.

We have run a large number of models in the
parameter space. First, we fitted
a single-planet model to the simulated observations and minimized the residuals.
In this step, we have run 1000 randomly simulated planet to map the structure of
the grid, to estimate the minimum. Then we refined the grid locally and run
$10^4$ planet templates. Among them, $\approx$ 10 possible planet solutions 
gave equivalently small residuals, ensuring that the grid was fine enough. 
The parameters of the best
fitting planet were input parameters for modelling the moon.

In the second step, 1.5$\times 10^6$ transits were simulated with 
random initial geometries of the satellite. The templates were allowed
to shift in time while fitting to the simulated observations. These resulted
in 2.7$\times 10^9$ fittings altogether.

The likelihood limit defining the 95\%{} 
confidence region has been deduced from 
the surface of  $rms$ scatter experimentally. 
For this, 1000 bootstrap observations were calculated using the
exact system parameters, noisified with different realizations 
of the same simulated observational noise. Then we determined the 
best-fit model planet, and calculated the residuals to the 
$M$ (simulation input) moon model.  
We mapped the 95\%{} acceptance regions of parameters, therefore the
$rms$ scatter limit was set to include 
95\%{} of these deviations. The resulting limits 
were $rms$ scatter of 0.149 m/s, 0.313 m/s and 0.609
m/s for the simulated observations of S/N=5, 2, 1 quality, respectively. These are, of course, particularly optimistic scenarios. 


\section{Discussion}

\begin{figure}
\includegraphics[width=8cm]{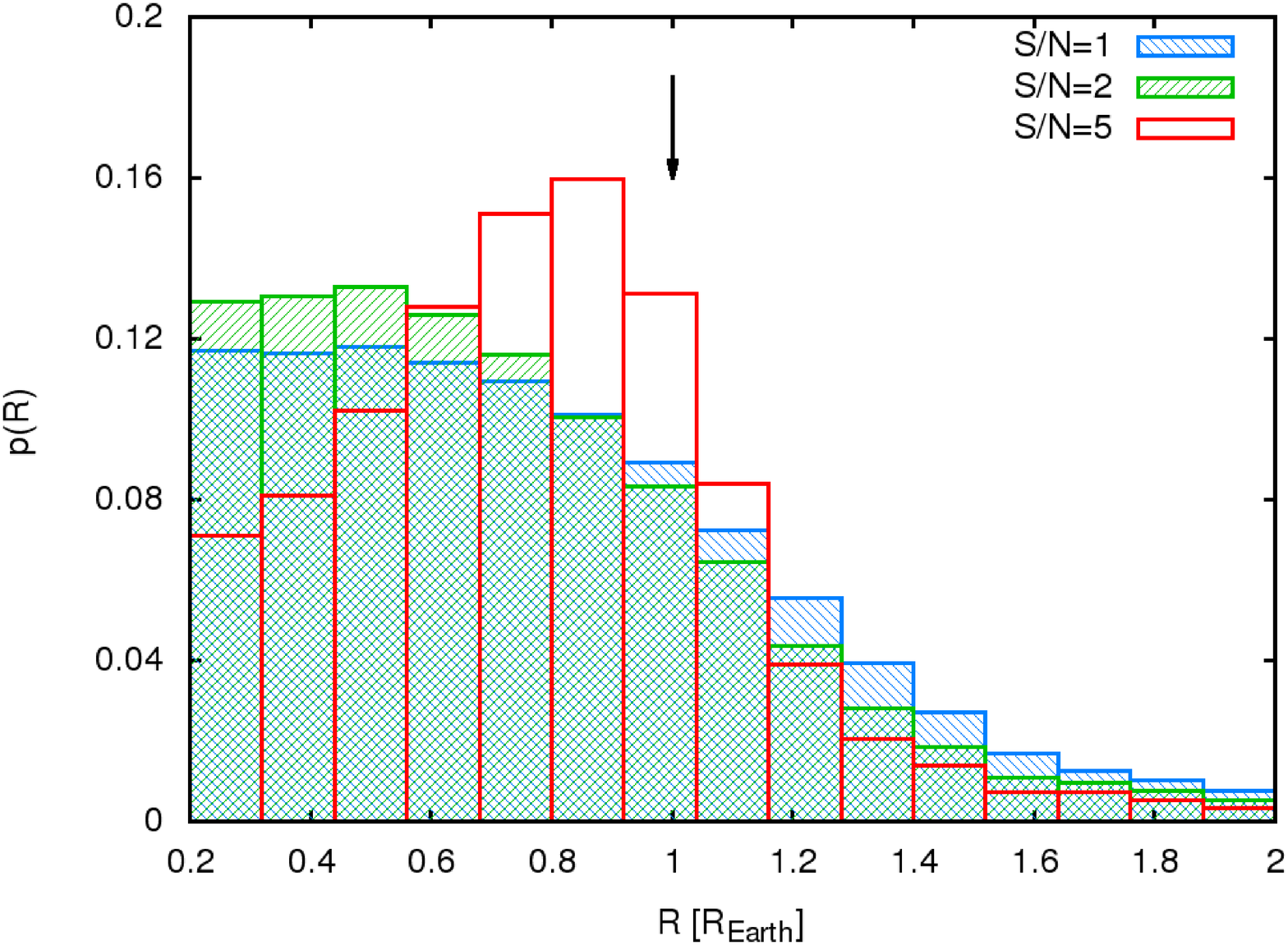}
\includegraphics[width=8cm]{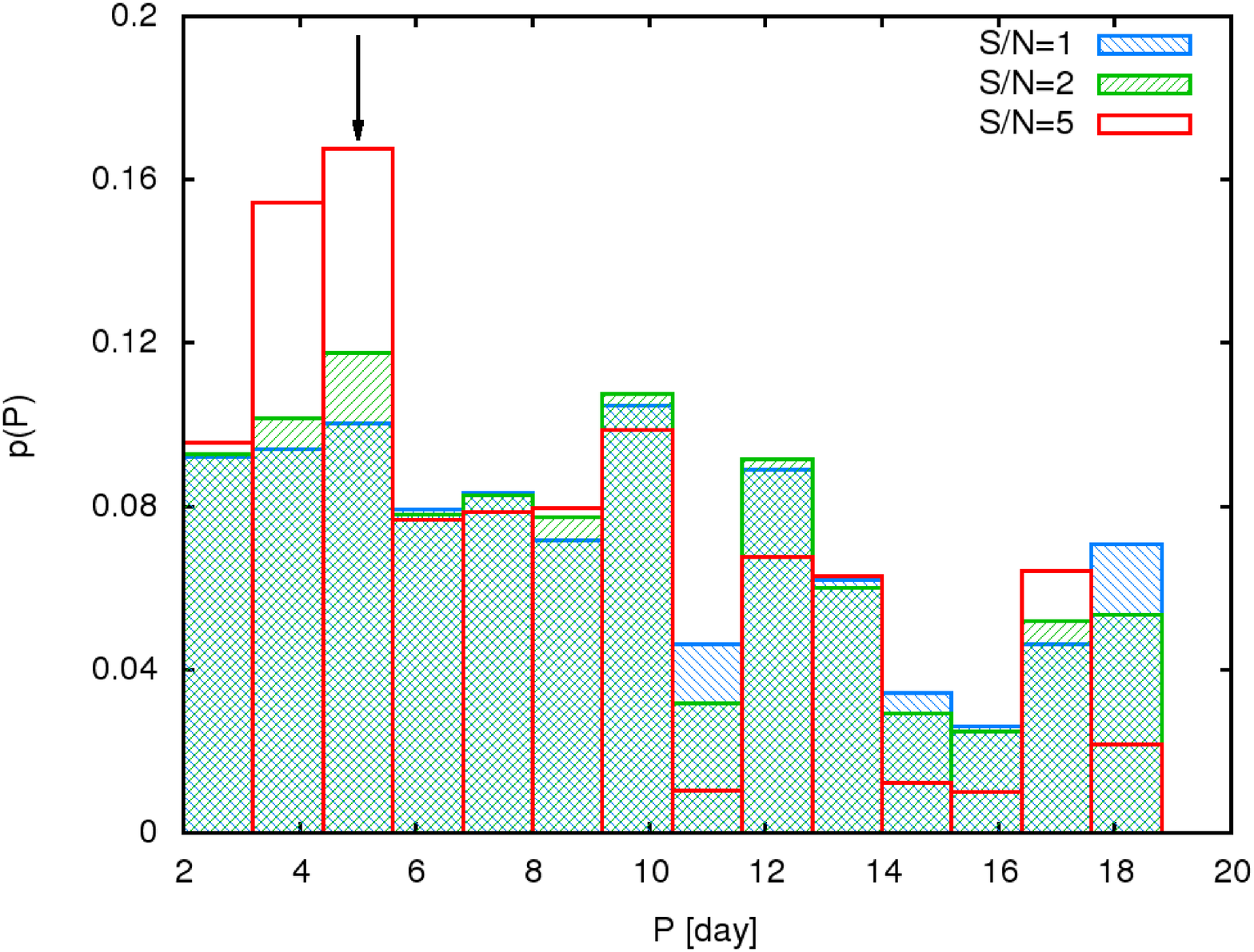}
\caption{Posterior probability distributions of $R$ and $P$,
marginalized from the fit likelihoods in the $P$--$R$ subspace
with uniform priors. The three different distributions refer to S/N
levels of 1, 2 and 5 with the same colour coding as in Figs. 3--4--5.
The arrow shows the input value of the moon parameters.}
\end{figure}

The joint confidence intervals of the parameter pairs are shown in 20 panels
in Figs. 4 and 5. Fig. 4 shows sections where the plotted parameters promise
a reliable reconstruction. The best results are given for the size
of the moon, that is closely linked to the amplitude of the residuals in the
RM curve, after subtracting the best-fit planet. The radius is
well reproduced in all sections and did not suffer serious degenerations. The topmost
row suggests that the size is a little biased and the inclination of the
moon is essentially unknown. This is because the residuals of the planet fitting is
forced to be close to zero, consequently the size of the moon is
slightly underestimated. For this smaller moon, a lower inclination parameter
is preferred. 

The ascending node of the orbit is somewhat correlated with the radius. Interestingly,
once other parameters are well constrained, $\Omega$ can be determined very accurately
as Fig. 4 shows (see the right panel in the second row). 
The 3rd and 5th rows of Fig. 4 demonstrate that there
is no information on the mass of the satellite from the analysis of radial
velocity, all probed masses are equally probable. The most important conclusion
is that the mass of the satellite must be omitted from the analysis, i.e. a fixed
value for the mass or a fixed assumption on the density of the satellite will
lead to equally reliable solutions in the other parameters with a much faster
process. The abscissa of the 4th and 5th panels in Fig. 4 show the orbital period of the moon.
Surprisingly, there is some information in the light curve for this period, as
values between 2--10 days are preferred (the model moon has 5 days period). 
Because of the little position change of the moon during the transit, the residual
RM effect due to the moon will last a somewhat shorter or longer 
time than that of the planet, which may be detected.

Fig. 5 shows further sections where degenerations are prominent. The joint
confidence intervals evidently show that the angle parameters,
$\iota$, $\phi$ and $\Omega$ are seriously interrelated.
Acceptable solutions can be characterized with very different values
of $\Omega$ or $\phi$, because the orbital period of the moon is unknown (see the 2nd and 3rd
rows). The 4th row shows that $\iota$ data do not constrain $\phi$ well. 
We suggest that a reliable value for $\iota$ 
will have to be assumed in practical applications, and
the other two angle parameters should be taken out of fitting (they can be 
either marginalized
or evaluated along some prior with Bayesian analysis).

However, these limitations of parameter determination do not lessen dramatically
the power
of radial velocity analysis in the exploration of exomoons. This method
nicely completes the evaluation of photometry, since the mass can be determined
from (barycentric) TTV. Simon et al. (2007) showed that
from photocentric TTV$_{\rm p}$ we get some information on the radius of the satellite,
but this is correlated with the density of the moon. The analysis of RM effects
give prior information on the size of the moon, and the combination of all methods,
theoretically, may lead to the direct experimental determination of the density
of the exomoons.

In Fig. 6, we show the posterior probability distributions of $P$ and $R$,
marginalized from the likelihood data with assuming uniform priors to all 
variables. The reconstruction of $R$ is satisfactory, although the radius
is somewhat biased toward smaller sizes. There is some information
for the period, too, which is somewhat surprising as the orbital period of
the moon is $\approx 1/12$ transit duration. The increasing noise level
does not bias the mode of the distributions, but gives the wings slightly
more weight.

\subsection{The general case of main sequence dwarf stars}

We have shown the reconstruction of parameters for a 0.8 M$_\odot$ main-sequence star,
which has a spectral type of K0 in case of solar metallicity. 
It is very important to note that our results are general and indicative for other types of stars. Although the signal will be smaller for bigger stars, the shape of the curves do not change significantly, hence error propagation will follow the same scenario as presented in Figs. 4-5. Consequently, the general behaviour of the parameters concerning their stability and degenerations will be the same. For a general case, we propose that the moon's radius is the best parameter for reconstruction; in some cases, the orbital period might be also constrained.

Earlier type stars have larger radius, hence the area eclipsed by the 
moon is a smaller fraction of the projected stellar disk. The other parameter
that determines the $A_{RM}$
half-amplitude of the RM effect is the $v_{rot} \sin i$
rotation velocity of the star, such that
\begin{equation}
A_{RM} \ \propto \ {\left( R \over R_*\right)^2 } \ v_{rot}\sin i,
\end{equation}
where $R$ denotes the radius of the moon, $R_*$ is the stellar radius and we can assume that $\sin i \approx 1$ for those systems that display transits and RM effect. Equality is true in Eq. 4 if we neglect limb darkening. Limb darkening can reduce the amplitude of the RM effect by 20--40\%, depending on the exact intensity profile of the stellar disk.
We can combine the RM effect of a moon and a planet together,
\begin{equation}
A_{RM,m+p} \propto {R_p^2 + R^2 \over R_*^2} \ v_{rot}.
\end{equation}
Now an upper estimate can be given
for the size of the moon if it is not detected in the residuals of the RM curve.
In this case the RM effect of the moon is hidden in the scatter of the 
radial velocity data, i.e. $3 \sigma_{v_{rad}}$ is larger than the satellite's effect:
\begin{equation}
3 \sigma_{v_{rad}}>A_{RM}={R^2\over R_*^2}v_{rot}.
\end{equation}
Rearranging this for the size of the moon, the upper limit is given as
$R<\sqrt{3 \sigma_{v_{rad}} / v_{rot}} R_*$, or simply substituting the amplitude of the measured RM effect (and assuming $A_{RM, m+p} \approx A_{RM,planet} = R_p^2/R_*^2 v_{rot}$),
\begin{equation}
R<\sqrt{3 \sigma_{v_{rad}} \over A_{RM, planet}} R_p.
\end{equation}

The confidence of this estimate is 99.9 \%, i.e. $3 \sigma$ confidence.

Which spectral types represent the best candidates for a successful detection of exomoons with the RM effect? To answer this question, we performed simple calculations for stars in the Pleiades open cluster, assuming that every star has a planet with a Ganymede-sized moon in central transit. Our intention was to predict the amplitude of the satellite's RM effect (assumed to be independent of that of the planet), as a function of stellar mass.

The $B-V$ colours and $v\sin i$ data for Pleiades stars were taken from Queloz et al. (1998).
Stellar masses and radii were estimated from the $B-V$ colour, using the latest Padova isochrones (Bertelli et al. 2008), adopting 70 Myr for the age and Z=0.017 for the metallicity (Boesgaard \& Friel 1990). We calculated the amplitude of the RM effect according to Eq. 4, inserting the stellar radius and rotation velocity for each star, and substituting the size of Ganymede. The results are plotted in Fig. 7 with open circles.

Although there are hints of a tendency, the scatter is large, which can be explained by the different spin axis orientations and different rotation evolution for each star. The shape of the distribution in Fig. 7 can be better seen using a statistical relationship for the rotation period of main-sequence stars determined by Barnes (2007): $P_{rot} \propto (B-V-0.4)^{0.601} t^{0.52}\ $ days, where $t$ is the age of the star. This formula is singular at $B-V=0.4$, thus it is valid for stars later than F5, approximately. Similarly to the individual stars, the $B-V$ colours of the isochrone points have been converted to masses and radii. The rotation velocity has then been calculated as $v_{rot}=2\pi R_*/P$ (we kept assuming $\sin i\approx 1$). To estimate RM amplitudes, we again used Eq. 4 and the size of Ganymede. In Fig. 7, the solid line shows the resulting average RM amplitude. We conclude that stars below 0.6--0.8 M$_\odot$ offer the best opportunity to detect the RM effect of the exomoons. Stars with masses greater than 1.2 M$_\odot$ do also show larger effect, however, stellar variability quickly becomes dominant with the increasing mass.

\begin{figure}
\includegraphics[angle=270,bb= 153 100 534 580,width=8cm]{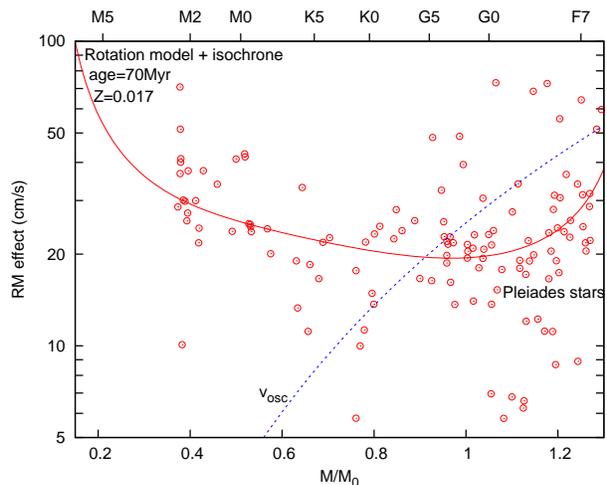}
\caption{Half amplitude of the Rossiter-McLaughlin effect due to Ganymede-sized
moons of planets orbiting G, K and M dwarfs. Solid line: empirical model 
based on Barnes (2007) rotation model and isochrones. Open
circles: individual stars of the Pleiades. The dashed curve shows the 
amplitude of solar-like oscillations.}
\end{figure}

This variability has two dominant components: the jitter due to convective motions 
(including the convectively excited solar-like oscillations) and the rotational modulation due to stellar activity.
Among the brighter dwarf stars, old, inactive G and K dwarfs offer 
the best performance: a few stars are known to have $<$1 m/s jitter, while they typically
have jitter levels in the 1--5 m/s regime (Saar et al. 2003, 
Wright 2005, O'Toole et al. 2008). On the contrary, some F-type
stars display large jitter that even challenges asteroseismology
(see the example of Procyon in Arentoft et al. 2008) and the detection of
long-period planets (Lagrange et al. 2009).
The jitter from solar-like oscillations can be estimated via the scaling relation of the velocity amplitude, which depends on the light-to-mass ratio of the star:
\begin{equation}
v_{osc}={L/L_\odot \over M/M_\odot }\left( 23.4\pm 1.4 \right) \ \ {\rm cm/s}
\end{equation}
(Kjeldsen \& Bedding 1995). The predicted oscillation velocity amplitudes are also plotted in Fig. 7 with the dashed line (blue in colour). Since M-dwarf stars have very small $L/M$ ratio, and consequently the amplitude of solar-like oscillations is tiny, they are promising candidates to be quested for exomoons. These stars are faint in the visual, but recent work of  Bean et al. (2009), for example, has opened the door to the sub-m/s velocimetric accuracy in the infrared with CRIRES on VLT. 

It is worth noting that late M-dwarfs (beyond M4) exhibit larger rotation velocities (Jenkins et al. 2009), which makes them difficult targets for high-precision RV measurements because rapid rotation washes out the spectral features (Bouchy et al. 2001). Hence, the best targets for exomoon exploration are the K and early M dwarf stars, for which both rotation and activity reach a minimum level (Jenkins et al. 2009, Wright 2005).

In case of higher stellar activity, the situation is not entirely hopeless, as illustrated by a recent study by Queloz et al. (2009), who filtered out activity with a Fourier polynomial using the first rotation harmonics. This way they pushed the residuals from $\pm 20$ m/s to $\pm$ 5 m/s. Similar residual levels in fitting the RM effect were reached by Triaud et al. (2009).

It is known that the frequency of giant planets increases linearly 
with the parent-star mass for stars between 0.4 and 3 $M_{\odot }$ 
(Ida \&{} Lin 2005, Kennedy \&{} Kenyon 2008), with e.g., 6$\%$ frequency of giant 
planets around 1 $M_{\odot }$ and 10$\%$ frequency around 1.5 $M_{\odot }$.
However, we know planets around red dwarfs and there is observational indication
for a few multiple planetary systems among them (e.g. Rivera et al. 2005).
The possible detection of exomoons around planets of larger stars, 
if these satellites ever exist, is a significant challenge for signal processing to minimise the ambiguity
caused by the higher level of velocity jitter.

A more elaborated distinction between signals of exoplanets and signals from stellar physics requires a deep analysis that is beyond the scope of this paper. There is reason for some optimism because the time-scales of the exoplanet-exomoon systems and that of the stellar signals are usually very different. Moreover, rapid development in instrumentation may reach levels of precision that were unimaginable even a few years ago.

\section*{Acknowledgments}

This work has been supported by the ``Lend\"ulet''
Young Researchers Program, the Bolyai J\'anos
Research Fellowship of the Hungarian Academy of 
Sciences, and the Hungarian OTKA Grants K76816 and K68626.


\end{document}